\documentclass[a4paper,fleqn,usenatbib]{mnras}

\usepackage{newtxtext,newtxmath}

\UseRawInputEncoding

\usepackage[T1]{fontenc}
\usepackage{ae,aecompl}
\usepackage{pdflscape}
\usepackage{rotating}
\usepackage{tablefootnote}
\usepackage{longtable}
\usepackage{threeparttable}
\usepackage{caption}
\usepackage{comment}

\usepackage{graphicx}	
\usepackage{amsmath}	
\usepackage{amssymb}	
\usepackage[LGRgreek]{mathastext}

\title[HDCO and D$_{2}$CO in proto-BDs]{A survey of HDCO and D$_{2}$CO towards Class 0/I proto-brown dwarfs}

\author[Riaz \& Thi]{
Riaz, B.$^{1}$\thanks{E-mail: briaz@usm.lmu.de}
\& Thi, W.-F.$^{2}$
\\
$^{1}$  Universit\"{a}ts-Sternwarte M\"{u}nchen, Ludwig Maximilians Universit\"{a}t, Scheinerstra$\beta$e 1, 81679 M\"{u}nchen, Germany
\\
$^{2}$  Max-Planck-Institut f\"{u}r Extraterrestrische Physik, Giessenbachstrasse 1, 85748 Garching, Germany
\\
}

\date{Accepted XXX. Received YYY; in original form ZZZ}

\pubyear{2022}

\begin{document}
\label{firstpage}
\pagerange{\pageref{firstpage}--\pageref{lastpage}}
\maketitle

\begin{abstract}

Deuterium fractionation can constrain the physical and chemical conditions at the early stage of brown dwarf formation. We present IRAM 30m observations over a wide frequency range of 213-279 GHz of singly and doubly deuterated species of formaldehyde (HDCO and D$_{2}$CO) towards Class 0/I proto-brown dwarfs (proto-BDs). Multiple low-excitation HDCO and D$_{2}$CO transition lines with upper energy level $\leq$ 40 K are detected. The D$_{2}$CO/HDCO, HDCO/H$_{2}$CO, D$_{2}$CO/H$_{2}$CO abundance ratios range between 0.01 and 2.5 for the proto-BDs, similar to the range seen in low-mass protostars. The highest ratios of D$_{2}$CO/HDCO $\sim$1.3-2.5 are measured for two Stage 0 proto-BDs. These objects could possess a warm corino, similar to the few hot corino cases reported among Class 0 protostars. The mean D$_{2}$CO/HDCO, D$_{2}$CO/H$_{2}$CO, and HDCO/H$_{2}$CO ratios for the proto-BDs are comparatively higher than the range predicted by the current gas-grain chemical models, indicating that HDCO and D$_{2}$CO are formed via grain surface reactions in the dense and cold interiors of the proto-BDs at an early formation stage.

\end{abstract}


\section{Introduction}
\label{intro}

Deuterium fractionation is a common characteristic of dark molecular clouds and star-forming region chemistry. In this context, formaldehyde has been an important molecule to study the physical conditions required to form highly deuterated molecules. Formaldehyde is an intermediate in the formation pathway to complex organic molecules. The singly and doubly deuterated forms of formaldehyde (HDCO and D$_{2}$CO) can form via gas-phase or grain surface reactions. 
In the gas-phase, deuterated formaldehyde primarily forms via the reaction chain (Turner 2001; Roueff et al. 2007):
\begin{gather}
CH_2D^{+} \xrightarrow{H_2} CH_{4}D^+ \xrightarrow{e^-} CH_{2}D \xrightarrow{O} HDCO,	\\
CHD^+_{2} \xrightarrow{H_2} CH_{3}D_{2}^+ \xrightarrow{e^-} CHD_{2} \xrightarrow{O} D_{2}CO.
\end{gather}
The reaction path starts with C$^+$ to form CH$_3^+$ and requires atomic oxygen to be present. This path is most efficient in the temperature range 30-100~K.

The formation on dust grains is through the multiple hydrogen addition reactions with frozen CO in a cold phase (e.g., Tielens 1983;  Watanabe \& Kouchi 2002; Hidaka et al. 2009; Fuchs et al. 2009):
\begin{gather}
CO \xrightarrow{H/D} HCO /HDO \xrightarrow{H/D} H_{2}CO / HDCO / D_2CO,   
\end{gather}
where all the species are ice species, followed by desorption in a warmer phase. The first step involves an energy barrier.
Taquet et al. (2012) suggest that deuterium substitution on grain surfaces can further enhance the deuteration:
\begin{gather}
H_{2}CO + \xrightarrow{D} HDCO + H \\
HDCO + D \xrightarrow{D} D_{2}CO + H.
\end{gather}
The reverse reactions only occur at high temperatures because the zero-point energy decreases when the deuteration increases.

We present the first observations of HDCO and D$_{2}$CO conducted with the IRAM 30-m telescope towards early-stage Class 0/I proto-brown dwarfs (proto-BDs). This work is a continuation of a survey of deuterated molecules towards proto-BDs in different star-forming regions and to derive the molecular D/H ratios from various species for these objects. The first results from our survey were presented in Riaz \& Thi (2022a), wherein we reported a wide range in the molecular D/H ratios of $\sim$0.02-3 derived from the DCO${+}$, DCN, DNC, and N$_{2}$D$^{+}$ species. No correlation was seen between the D/H ratios and the CO depletion factor or the evolutionary stage of the system. We found trends of higher DCO$^{+}$/HCO$^{+}$ and DCN/HCN ratios, nearly constant DNC/HNC and DNC/HN$^{13}$C ratios, but lower N$_{2}$D$^{+}$/N$_{2}$H$^{+}$ ratios in the proto-BDs compared to low-mass protostars. In addition, we reported the first detection in mono-deuterated methane (CH$_{3}$D) in three proto-BDs, with fractional abundances comparable to low-mass protostars (Riaz \& Thi 2022b). Both low-temperature gas-phase ion-molecule deuteron transfer and grain surface reactions are required to explain the enhanced deuterium fractionation. The very dense and cold (n$_{H_{2}}\geq$10$^{6}$ cm$^{-3}$, T $\leq$10 K) interior provide the suitable conditions for efficient deuterium fractionation in proto-BDs.

In the present work, we look at the D/H ratios derived from the singly and doubly deuterated species of formaldehyde, and investigate if deuteration increases towards more complex molecules. How can we relate the D/H ratios found in molecules to the overall D/H in brown dwarf atmosphere? It depends on whether the molecules are destroyed and reformed in the brown dwarf atmosphere, i.e. the lifetime of a given species. The HDCO/D$_{2}$CO ratio cannot change the HD/H$_{2}$ ratio in brown dwarf atmosphere but if one can measure HDCO/D$_{2}$CO, there could be a relation with that same value in the brown dwarf environment. In fact, simple species like HCN/DCN may survive quite a long time even after incorporation into the atmosphere (for e.g., the case of Shoemaker-Levy 9 comet). Upon incorporation in the proto-BD atmosphere, simple organic molecules may survive long enough till the brown dwarf phase and be detected. The D/H ratio in those molecules may also be preserved. However, those ratios are most likely different than the HD/H$_{2}$ ratio.

Section~\ref{obs} describes the sample of proto-BDs studied in this work, the IRAM 30-m observations, and data reduction. The measurements of the HDCO and D$_{2}$CO column densities and molecular abundances are presented in Sect.~\ref{analysis}. A discussion on the molecular D/H ratios, their dependence on the CO depletion factor, a comparison with low-mass protostars and predictions from chemical models is presented in Sect.~\ref{results}.

\section{Sample, Observations, Data Reduction}
\label{obs}

\begin{table*}
\centering
\caption{Sample}
\label{sample}
\begin{threeparttable}
\begin{tabular}{lcccccc} 
\hline
Object & L$_{bol}$ (L$_{\sun}$)\tnote{a} & Classification\tnote{b} & Region \\
\hline
SSTc2d J182854.9+001833 (J182854) &  0.05 &  Stage 0+I, Stage I, Class 0/I & Serpens \\ 
SSTc2d J182844.8+005126 (J182844) &  0.04 &  Stage 0+I, Stage 0, Class 0/I & Serpens \\ 
SSTc2d J182959.4+011041 (J182959) &  0.008 &  Stage 0+I, Stage I, Class 0/I & Serpens \\ 
SSTc2d J182856.6+003008 (J182856) & 0.004 &  Stage 0+I, Stage 0, Class 0/I & Serpens \\ 
SSTc2d J182952.2+011559 (J182952) & 0.024 &  Stage 0+I, Stage 0, Class 0/I & Serpens \\ 
SSTc2d J162625.62-242428.9 (J162625) & 0.04 &  Stage 0+I, Stage 0, Class 0/I & Ophiuchus \\  
SSTc2d J032838.78+311806.6 (J032838) & 0.017 &  Stage 0+I, Stage I, Class 0/I & Perseus \\   
SSTc2d J032848.77+311608.8 (J032848) & 0.013 &  Stage 0+I, Stage I, Class 0/I & Perseus \\   
SSTc2d J032851.26+311739.3 (J032851) & 0.011 &  Stage 0+I, Stage 0, Class 0/I & Perseus \\  
SSTc2d J032859.23+312032.5 (J032859) & 0.006 &  Stage 0+I, Stage I, Class 0/I & Perseus \\   
\hline
\end{tabular}
\begin{tablenotes}
\item[a] Errors on L$_{bol}$ is estimated to be $\sim$20\%. 
\item[b] The first, second, and third values are using the classification criteria based on the integrated intensity in the HCO$^{+}$ (3-2) line, the physical characteristics, and the SED slope, respectively.
\end{tablenotes}
\end{threeparttable}
\end{table*}

We observed several HDCO and D$_{2}$CO lines in the 213 -- 279 GHz frequency range towards the complete sample of 16 Class 0/I proto-BDs described in Riaz \& Thi (2022a). Out of these 16 targets, 10 proto-BDs show emission in HDCO and/or D$_{2}$CO. Table~\ref{sample} lists the properties for these 10 proto-BDs. A detailed discussion on the target selection, classification and measurements of mass, luminosity, H$_{2}$ number and column density for these objects can be found in Riaz et al. (2018) and Riaz \& Thi (2022a).

The observations were obtained at the IRAM 30-m telescope between 2017 and 2021. We used the EMIR heterodyne receiver (E230 band), and the FTS backend in the wide mode, with a spectral resolution of 200 kHz ($\sim$0.3 km s$^{-1}$ at 232 GHz). The observations were taken in the frequency switching mode with a frequency throw of approximately 7 MHz. The source integration times ranged from 3 to 4 hours per source per tuning reaching a typical RMS (on {\it T}$_{A}^{*}$ scale) of $\sim$0.01-0.02 K. The telescope absolute RMS pointing accuracy is better than 3$\arcsec$ (Greve et al. 1996). All observations were taken under good weather conditions (0.08 $<$ $\tau$ $<$0.12; PWV $<$2.5 mm). The absolute calibration accuracy for the EMIR receiver is around 10\% (Carter et al. 2012). The telescope intensity scale was converted into the main beam brightness temperature ({\it T}$_{mb}$) using standard beam efficiency of $\sim$59\% at 230 GHz. The half power beam width of the telescope beam is $\sim$10$\arcsec$ at 230 GHz. The spectral reduction was conducted using the CLASS software (Hily-Blant et al. 2005) of the GILDAS facility\footnote{http://www.iram.fr/IRAMFR/GILDAS}. The standard data reduction process consisted of averaging multiple observations, extracting a subset around the line rest frequency, and fitting a low-order polynomial baseline which was then subtracted from the average spectrum. Observations at offset positions were carried-out. Emissions are only seen at the location of the sources.

Table~\ref{lines} provides details of the HDCO and D$_{2}$CO transitions lines detected towards our targets. At present, reaching the same level of sensitivity of $\sim$10-20 mK, HDCO is detected in 6 proto-BDs, 4 of which also show emission in the D$_{2}$CO lines. In addition, there are 2 proto-BDs that show emission in D$_{2}$CO but no detection in HDCO; however the upper limit on the HDCO fluxes for these two objects are comparable to D$_{2}$CO. The inconsistency in the HDCO and D$_{2}$CO detection could be due to differences in the weather conditions since the two lines were not observed during the same night for all objects. Among the HDCO detections, 5 out of 6 show emission in the ortho-HDCO 4(0,4)-3(0,3) transition line, one of which (J032859) also shows emission in the para-HDCO 4(1,4)-3(1,3) line, while one object (J182856) only shows emission in the para-HDCO 4(1,3)-3(1,2) line. Among the D$_{2}$CO detections, all 6 proto-BDs show emission in the ortho-D$_{2}$CO 4(0,4)-3(0,3) transition line. In addition, J162625 shows emission in the para-D$_{2}$CO 4(1,4)-3(1,3) line and the para-D$_{2}$CO 4(1,3)-3(1,2) line is also detected in J032859. As can be seen from Table~\ref{lines}, most of the detections are in the transition line with the lowest upper energy level (E$_{upper} <$ 31 K) within our frequency range. None of the HDCO and D$_{2}$CO transition lines with E$_{upper} >$ 40 K between 213 and 279 GHz were detected for our proto-BD sample.

\begin{table*}
\centering
\caption{Molecular line observations}
\label{lines}
\begin{threeparttable}
\begin{tabular}{ccccccc} 
\hline
Molecule		& Transition	&	Frequency (GHz)	&	log (Intensity)	&	log(A$_{ij}$)	&	E$_{up}$ (K)	\\
\hline
p-HDCO		&	4(1,4)-3(1,3)	&	246.9246		& -2.42	&	-3.40		&	37.60		\\
o-HDCO		&	4(0,4)-3(0,3)	&	256.5855		& -2.35	&	-3.32		&	30.85		\\
p-HDCO		&	4(1,3)-3(1,2)	&	268.2920		& -2.35	&	-3.29		&	40.17		\\
\hline
p-D$_{2}$CO	& 	4(1,4)-3(1,3) 	& 	221.1916		& -2.79	& 	-3.54		& 	31.96	\\
o-D$_{2}$CO	& 	4(0,4)-3(0,3)	& 	231.4102		& -2.42	& 	-3.46		&	27.88	\\
p-D$_{2}$CO	& 	4(1,3)-3(1,2)	& 	245.5327		& -2.71	&	-3.41		& 	34.88	\\

\hline
\end{tabular}
\end{threeparttable}
\end{table*}

\section{Data Analysis}
\label{analysis}

The HDCO and D$_{2}$CO spectra are shown in Figs.~\ref{hdco-figs} and ~\ref{d2co-figs}, respectively. We have measured the parameters of the line center, line width, the peak and integrated intensities using a single- or double-peaked Gaussian fit. A double-peaked Gaussian may stem mostly from either low signal-to noise, Keplerian rotation, or self-absorption. The line parameters are listed in Table~\ref{hdco-pars} and ~\ref{d2co-pars}. The uncertainty is estimated to be $\sim$20\% for the peak and integrated intensities and $\Delta${\it v}, and $\sim$0.02-0.04 km/s for {\it V}$_{lsr}$. The errors on the line parameters are due to uncertainties in fitting the line profile, and mainly arise from the end points chosen for the Gaussian fit. For the non-detections, a 2-$\sigma$ upper limit was calculated by integrating over the velocity range of $\pm$2 km s$^{-1}$ from the cloud systemic velocity of $\sim$8 km s$^{-1}$ for Serpens and Perseus, and $\sim$4 km s$^{-1}$ for Ophiuchus (e.g., White et al. 2015; Burleigh et al. 2013).

The column density of HDCO and D$_{2}$CO is derived from the integrated line intensity by assuming local thermodynamic equilibrium (LTE) condition and an excitation temperature, T$_{ex}$, of 10 K, 30 K, and 50 K. The method is described in detail in Riaz \& Thi (2022b). The kinetic temperature is expected to be relatively low ($\sim$10 K) throughout the outer and inner envelope layers in the proto-BDs (e.g., Machida et al. 2009), with only the innermost ($<$20 au) region to be at higher temperatures ($>$ 20 K). This would explain the non-detection in the HDCO and D$_2$CO lines with $E_{\rm upper} >$ 40~K for our proto-BD sample.

The column densities and the molecular abundances relative to H$_{2}$, [{\it N}(X)/{\it N}(H$_{2}$)], are listed in Table~\ref{hdco-abund} and ~\ref{d2co-abund}. The column densities and fractional abundances are nearly constant with T$_{ex}$ and increase slightly by a factor of $\sim$1.5 for a few objects when T$_{ex}$ is increased from 10 K to 50 K. The error on the column densities is propagated from the error on the integrated line intensity, and is estimated to be $\sim$20\%. The error on the abundances is propagated from the error on the HDCO, D$_{2}$CO, and H$_{2}$ column densities, and is estimated to be $\sim$23\%-24\%.

\begin{table*}
\centering
\caption{HDCO Column Densities and Molecular Abundances}
\label{hdco-abund}
\begin{threeparttable}
\begin{tabular}{lcccccccccccccccc} 
\hline
Object & ${\it N}_{H_{2}}$  & \multicolumn{3}{c}{${\it N}$(HDCO) (x10$^{12}$ cm$^{-2}$) \tnote{a}} 	& \multicolumn{3}{c}{[HDCO] (x10$^{-12}$) \tnote{a}} 	\\
\hline
	   & (x10$^{22}$ cm$^{-2}$) & 10 K & 30 K & 50 K & 10 K & 30 K & 50 K  	\\
\hline
\multicolumn{8}{c}{o-HDCO 4(0,4)-3(0,3)}	\\ 
\hline 				     
J182854	& 3.7$\pm$0.6 		& <0.2 & <0.2 & <0.3 	 	& <4.5 & <4.6 & <7.0 	\\
J182844 	& 3.6$\pm$0.5 		& 0.2 & 0.2 & 0.3 			& 5.7 & 5.8 & 9.0 		\\
J182959	& 4.5$\pm$0.5		& <0.2 & <0.2 & <0.3 	 	& <3.7 & <3.7 & <5.8 	\\
J182856	& 3.4$\pm$0.4		& <0.2 & <0.2 & <0.3 	 	& <4.8 & <5.0 & <7.7 	\\
J182952	& 8.7$\pm$0.6		& 0.3 & 0.3 & 0.5 			& 3.3 & 3.4 & 5.3 		\\
J162625	& 4.0$\pm$0.6		& 0.8 & 0.9 & 1.4 	 		& 21.7 & 22.2 & 34.4 	\\
J032838	& 21.3$\pm$3.0	& 0.3 & 0.3 & 0.5 	 		& 1.4 & 1.4 & 2.1 	\\
J032848	& 10.4$\pm$1.5	& 0.2 & 0.2 & 0.3 	 		& 1.6 & 1.6 & 2.5 	\\
J032851	& 7.3$\pm$2.2		& <0.2 & <0.2 & <0.3 	 	& <2.3 & <2.3 & <3.6 	\\
J032859	& 34.2$\pm$12.4	& 0.6 & 0.7 & 1.0 			& 1.9 & 1.9 & 3.0 	\\
\hline
\multicolumn{8}{c}{p-HDCO 4(1,4)-3(1,3)}	\\ 
\hline 
J032859	& 34.2$\pm$12.4	& 1.6 & 1.1 & 1.5 			& 4.8 & 3.1 & 4.4 	\\
\hline
\multicolumn{8}{c}{p-HDCO 4(1,3)-3(1,2)}	\\ 
\hline 
J182856	& 3.4$\pm$0.4		& 0.3 & 0.2 & 0.2 			& 9.1 & 5.1 & 7.0 	\\
\hline
\end{tabular}
\begin{tablenotes}
\item[a] The uncertainty is estimated to be $\sim$20\% for the column densities and $\sim$23\%-24\% for the molecular abundances. 
\end{tablenotes}
\end{threeparttable}
\end{table*}

\begin{table*}
\centering
\caption{D$_{2}$CO Column Densities and Molecular Abundances}
\label{d2co-abund}
\begin{threeparttable}
\begin{tabular}{llccccccccc} 
\hline
Object & ${\it N}_{H_{2}}$  & \multicolumn{3}{c}{${\it N}$(D$_{2}$CO) (x10$^{12}$ cm$^{-2}$) \tnote{a}} 	& \multicolumn{3}{c}{[D$_{2}$CO] (x10$^{-12}$) \tnote{a}} 	\\
\hline
	   & (x10$^{22}$ cm$^{-2}$) & 10 K & 30 K & 50 K & 10 K & 30 K & 50 K 	\\
\hline
\multicolumn{8}{c}{o-D$_{2}$CO 4(0,4)-3(0,3)}	\\ 	
\hline 				     
J182854 	& 3.7$\pm$0.6  	& 0.2 & 0.3 & 0.5  		& 6.8 & 8.0 & 12.8 	\\
J182844 	& 3.6$\pm$0.5 		& 0.07 & 0.08 & 0.1  		& 2.0 & 2.4 & 3.7 		\\
J182959 	& 4.5$\pm$0.5 		& 0.25 & 0.3 & 0.5  		& 5.6 & 6.6 & 10.5	\\
J182856	& 3.4$\pm$0.4		& <0.1 & <0.2 & <0.3  	& <4.3 & <5.0 & <8.0 	\\
J182952	& 8.7$\pm$0.6		& 0.4 & 0.4 & 0.7  		& 4.2 & 4.8 & 7.8 		\\
J162625	& 4.0$\pm$0.6		& 1.8 & 2.1 & 3.4  		& 45.5 & 53.2 & 84.8		\\
J032838	& 21.3$\pm$3.0	& <0.1 & <0.2 & <0.3  	& <0.7 & <0.8 & <1.3 	\\
J032848	& 10.4$\pm$1.5	& <0.1 & <0.2 & <0.3  	& <1.4 & <1.6 & <2.6 	\\
J032851	& 7.3$\pm$2.2		& <0.1 & <0.1 & <0.2  	& <1.6 & <1.7 & <2.8 	\\
J032859	& 34.2$\pm$12.4	& 0.4 & 0.4 & 0.7  		& 1.1 & 1.2 & 2.0 		\\
\hline
\multicolumn{8}{c}{p-D$_{2}$CO 4(1,4)-3(1,3)}	\\ 	
\hline 
J162625	& 4.0$\pm$0.6		& 1.4 & 1.2 & 1.8 		& 34.5 & 30.1 & 45.3 	\\
\hline
\multicolumn{8}{c}{p-D$_{2}$CO 4(1,3)-3(1,2)}	\\	
\hline
J032859	& 34.2$\pm$12.4	& 0.7 & 0.5 & 0.8 		& 2.1 & 1.6 & 2.3 	\\
\hline
\end{tabular}
\begin{tablenotes}
\item[a] The uncertainty is estimated to be $\sim$20\% for the column densities and $\sim$23\%-24\% for the molecular abundances. 
\end{tablenotes}
\end{threeparttable}
\end{table*}

\section{Results and Discussion}
\label{results}

For a quantitative comparison of the correlations between the various parameters, we have calculated the Pearson correlation coefficient, {\it r}, from the best-fit to the observed data points, and the value for {\it r} is noted in each plot. The Pearson correlation coefficient is a measure of the covariance of the two variables divided by the product of their standard deviations. A value of $\pm$1 for {\it r} indicates a strong correlation or anti-correlation between the two parameters for a positive or negative slope of the best-fit, respectively.

\subsection{D/H ratios}

Table~\ref{ratios} lists the D$_{2}$CO/HDCO, HDCO/H$_{2}$CO, D$_{2}$CO/H$_{2}$CO ratios for the proto-BDs. These ratios have been derived using the o-HDCO 4(0,4)-3(0,3), o-D$_{2}$CO 4(0,4)-3(0,3), and o-H$_{2}$CO 3(1,2)-2(1,1) transition lines. Most detections are in these ortho lines of HDCO and D$_{2}$CO (Table~\ref{hdco-abund};~\ref{d2co-abund}). Table~\ref{co-h2co} lists the ortho-H$_{2}$CO abundances from Riaz et al. (2019). The highest values are measured for the D$_{2}$CO/HDCO ratios, with a mean value of $\sim$1.1-1.25. The lowest are D$_{2}$CO/H$_{2}$CO ratios, with a mean value of $\sim$0.05-0.1. These ratios are within the wide range of $\sim$0.005-7 in the D/H ratios derived from six different deuterated species for proto-BDs (Riaz \& Thi 2022a).

The D$_{2}$CO/HDCO ratios show a possible dependence on the evolutionary stage of the proto-BD, such that high ratios of $>$1 are only measured for the Stage 0 objects J182952 and J162625 (Table~\ref{ratios}). The deuteration tends to decrease with the evolutionary stage due to the gradual collapse of the external shells of the envelope that are less deuterated because of their lower density. No clear dependence, however, is seen between the HDCO/H$_{2}$CO and D$_{2}$CO/H$_{2}$CO ratios and the evolutionary stage. Considering that all of these cold, dense proto-BDs have kinetic temperature $\leq$10 K, the non-detection in HDCO and D$_{2}$CO could be due to warmer environment, for e.g., due to strong shock emission knots close to the driving source, or thermal heating. Since low temperatures are required for high levels of deuteration to occur, lower deuteration levels suggest warmer environment.

The relation between the abundance ratios of HDCO and D$_{2}$CO can provide an indication of whether these species have a grain surface or gas-phase formation pathway (Turner 1990; Charnley et al. 1997; Rodgers \& Charnley 2002). These works have defined the quantity {\it F} as
\begin{gather}
{\it F} = (HDCO/H_{2}CO)^{2}/(D_{2}CO/H_{2}CO),
\end{gather}
\noindent where {\it F} has values of 1.3-1.6 for gas-phase chemistry and {\it F} $\geq$1 for a grain surface origin, although grain surface formation over a long timescale could result in lower values for {\it F}. For our proto-BD targets, {\it F} $\sim$0.1-0.2 (ignoring the upper limits; Table~\ref{ratios}), which is within the range of 0.03 to 0.4 measured for low-mass protostars (e.g., Bacmann et al. 2003; Loinard et al. 2002; Parise et al. 2006; Roberts et al. 2002, 2007; Kang et al. 2015; Persson et al. 2018). These values for {\it F}, however, are inconclusive about the gas-phase versus grain surface formation of formaldehyde, and the distinction between these pathways becomes more complicated due to the substitution and abstraction with subsequent addition reactions on grain surfaces (Sect.~\ref{intro}).

Assuming equal transmission probabilities for the CO + H and CO + D formation channels on grain surfaces for formaldehyde and neglecting abstraction reactions (Sect.~\ref{intro}), Turner (1990) and Charnley et al. (1997) estimated the D$_{2}$CO/HDCO ratios to be one-fourth of the HDCO/H$_{2}$CO ratios. For our proto-BD targets, ignoring the upper limits, the D$_{2}$CO/HDCO ratios are higher than the HDCO/H$_{2}$CO ratios (Table~\ref{ratios}). This highlights the importance of H and D abstractions and substitutions on formaldehyde that are as efficient as addition reactions, and could enhance the deuterium fractionation of H$_{2}$CO via grain surface reactions once CO is depleted (e.g., Taquet et al. 2012; 2014).

The large molecular D/H ratios measured in this work indicate that HDCO and D$_{2}$CO likely form via grain surface reactions in the dense and cold interiors of the proto-BDs, since gas-phase reactions are unable to reproduce the large observed abundances. Most notably for the case of the Stage 0 objects J182952 and J162625 with D$_{2}$CO/HDCO ratio $>$1, the high deuterium fractionation was likely set during the colder early formation phases when material was frozen onto grain surfaces. Surface reactions are extremely efficient and faster than gas-phase reactions since accretion onto grains can expedite the reactions. Later, during the protostellar phase, when the circumstellar envelope became warmer either due to thermal heating from the central proto-BD or some non-thermal desorption mechanism, the grain mantles evaporated resulting in high D/H ratios measured in the warm gas. Due to this, the gaseous D/H ratios strongly depend on the ratios in the ice mantles.

The two Stage 0 proto-BDs, J182952 and J162625, could be the sub-stellar analogs of the few hot corino cases reported among Class 0 protostars (e.g., IRAS 16293-2422 and NGC 1333 IRAS2A), albeit the temperature in the innermost, densest regions in proto-BDs is unlikely to be $>$50-70 K, lower than $\geq$100 K theorised for hot corino protostars (e.g., Ceccarelli et al. 1998). Grains accreted at small radii close to the central proto-BD are more likely to lose their ice coatings and released in the gas. Considering that the desorption temperature of HDCO and D$_{2}$CO is $\sim$37 K, the warm corino region close to the central proto-BD provides the right physical conditions for thermal desorption of HDCO and D$_{2}$CO formed on grain surfaces, as also noted to explain the detection of CH$_{3}$D and the possibility of warm carbon chain chemistry in proto-BDs (Riaz \& Thi 2022b).

\begin{table*}
\centering
\caption{D/H ratios}
\label{ratios}
\begin{threeparttable}
\begin{tabular}{lccccccccccccc} 
\hline
Object & \multicolumn{3}{c}{[o-D$_{2}$CO/o-HDCO] \tnote{a}} & \multicolumn{3}{c}{[o-HDCO/o-H$_{2}$CO] \tnote{a}} & \multicolumn{3}{c}{[o-D$_{2}$CO/o-H$_{2}$CO] \tnote{a}} 	\\
\hline
				& 10 K & 30 K & 50 K & 10 K & 30 K & 50 K  & 10 K & 30 K & 50 K 	\\
\hline		        
J182854 (Stage I)	& <1.5 & <1.8 & <1.8 	& <0.4 & <0.7 & <1.2   	 	& <0.7 & <1.3 & <2.2 	\\
J182844 (Stage 0)	& 0.3 & 0.4 & 0.4 	 	& 0.06 & 0.1   & 0.2   	   	& 0.02 & 0.04 & 0.07   	\\
J182959 (Stage I)	& <1.5 & <1.8 & <1.8 	& <0.01 & <0.02 & <0.04 	 	& 0.01 & 0.04 & 0.08  		\\
J182856 (Stage 0) 	& <0.9 & <1.0 & <1.0 	& <0.01 & <0.04 & <0.06 		& <0.01 & <0.04 & <0.06 	\\
J182952 (Stage 0)	& 1.3 & 1.4 & 1.5 	 	& 0.2	   & 0.1   & 0.1   	   	& 0.2   & 0.1   & 0.2   	\\
J162625 (Stage 0)	& 2.1 & 2.4 & 2.5 	 	& --	   & --      & --     	     	& --     & --      & --      	\\
J032838 (Stage I)	& <0.5 & <0.6 & <0.6 	& 0.1& 0.2   & 0.2   		 	& <0.06   & <0.1   & <0.1 	\\
J032848 (Stage I)	& <0.8 & <1.0 & <1.0 	& 0.07 & 0.1   & 0.2   	   	& <0.06   & <0.1   & <0.2  	\\
J032851 (Stage 0)	& <0.7 & <0.7 & <0.7 	& <0.03 & <0.06 & <0.08 		& <0.02 & <0.04 & <0.06	\\
J032859 (Stage I)	& 0.6 & 0.6 & 0.6	 	& 0.06 & 0.04 & 0.05 	 	& 0.04 & 0.02 & 0.03 	\\
\hline
Average 		& 1.07  & 1.2 & 1.25 & 0.098  & 0.108 & 0.15 & 0.074 & 0.051  & 0.098  \\
STD			& 0.7 & 0.78 & 0.83 & 0.053 & 0.051 & 0.063 & 0.08 & 0.032 & 0.067	\\
\hline
\end{tabular}
\begin{tablenotes}
\item[a] The uncertainty is estimated to be $\sim$23-32\% on the D/H ratios. The mean ratios and standard deviation (STD) are calculated excluding the upper limits. A `--' indicates that observations are not available. 
\end{tablenotes}
\end{threeparttable}
\end{table*}

\subsection{Dependence on CO depletion}

\begin{figure}
  \centering           
     \includegraphics[width=2.8in]{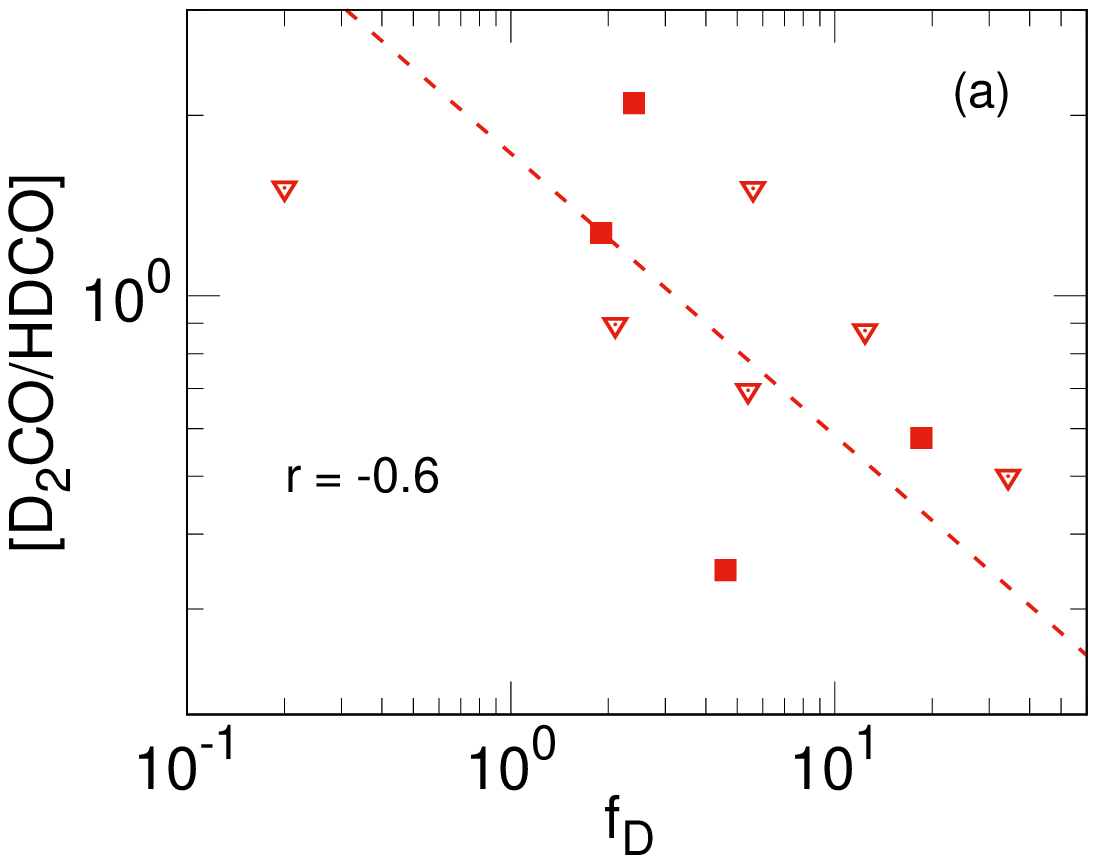}	 
     \includegraphics[width=2.8in]{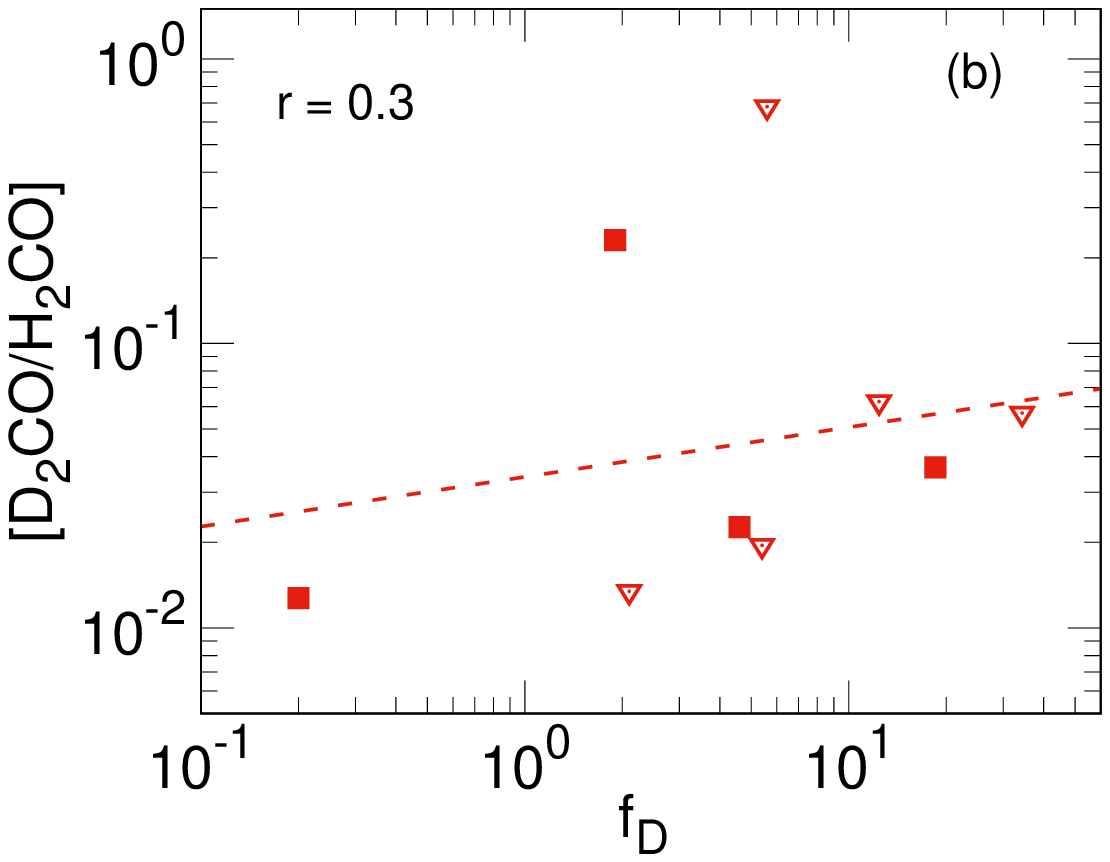}
     \includegraphics[width=2.8in]{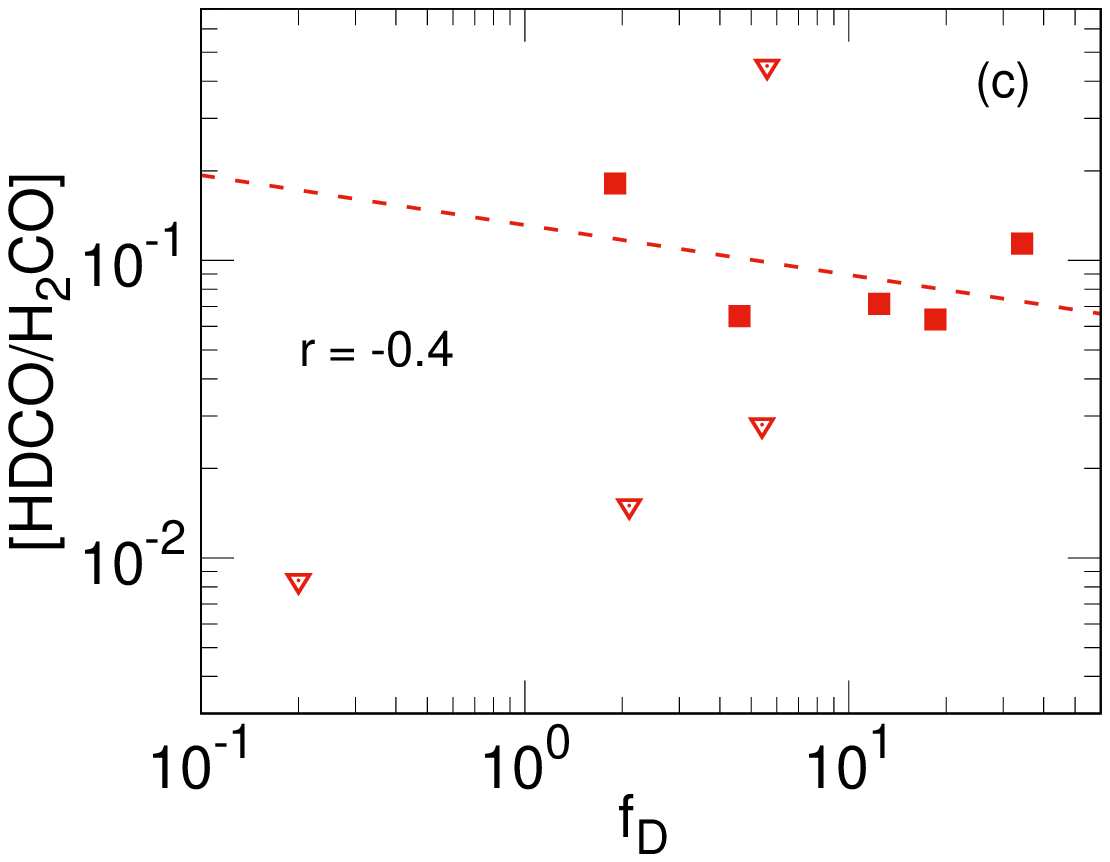}
     \caption{Correlations of the D$_{2}$CO/HDCO ({\bf a}), D$_{2}$CO/H$_{2}$CO ({\bf b}), and HDCO/H$_{2}$CO ({\bf c}) ratios with the CO depletion factor, $f_{D}$. Diamonds denote the non-detections (upper limits). Dashed line is the best-fit to the observed data points excluding the upper limits. The correlation coefficient for the best-fit is noted in the plot. }
     \label{co-plots}     
  \end{figure}

Figure~\ref{co-plots} probe the dependence of the D$_{2}$CO/HDCO, D$_{2}$CO/H$_{2}$CO, and HDCO/H$_{2}$CO ratios on the CO depletion factor. The CO depletion factor, $f_{D}$, is defined as $X_{can}/X$ (Bacmann et al. 2002), where $X_{can}$ is the canonical abundance determined by Frerking et al. (1982) towards dark cores ($X_{can}$ = 4.8 $\times$ 10$^{-8}$), and the abundance $X$ is the ratio of the C$^{17}$O and H$_{2}$ column densities. For $f_{D} <$1, CO is expected to be in the gas-phase, while higher values of $f_{D} \sim$10 indicate a large fraction ($>$80\%) of CO is frozen onto dust grains. 

The D$_{2}$CO/HDCO ratios show a strong anti-correlation ({\it r} = -0.6) with $f_{D}$ (Fig.~\ref{co-plots}a), suggesting that deuteration tends to be more efficient at low CO depletion levels. No strong correlation, however, is seen between the D$_{2}$CO/H$_{2}$CO and HDCO/H$_{2}$CO ratios and CO depletion factor (Fig.~\ref{co-plots}bc), suggesting that deuteration of H$_{2}$CO is active in all regions of the proto-BDs. If the main formation pathway of HDCO and D$_{2}$CO is via surface grain chemistry (Sect.~\ref{intro}) then deuteration in proto-BDs is likely most efficient in a warm narrow region in the envelopes where CO is not severely depleted and the abundance is high enough (10$^{-4}$ - 10$^{-5}$) for the reaction to proceed (e.g., Riaz \& Thi 2022a).

\begin{table}
\centering
\caption{ortho-H$_{2}$CO Molecular Abundances}
\label{co-h2co}
\begin{threeparttable}
\begin{tabular}{lccccccc} 
\hline
Object & $f_{D}$ & \multicolumn{3}{c}{[o-H$_{2}$CO] (x10$^{-12}$) \tnote{a}}   \\ 
\hline
	   &		   &    10 K    	& 30 K 	& 50 K 	\\
\hline
J182854 	& 5.6		& <10.0	& <6.0	& <5.7	\\
J182844 	& 4.6		& 88.4	& 58.3	& 51.4	\\
J182959 	& 0.2		& 440.0	& 146.0	& 131.0	\\
J182856	& 2.1		& 320.0	& 126.0	& 123.0	\\
J182952	& 1.9		& 18.2	& 36.7	& 47.1	\\
J162625	& 2.4		& --		& --		& --		\\
J032838	& 34.3	& 12.3	& 7.4		& 9.1		\\
J032848	& 12.4	& 22.4	& 12.5	& 14.4	\\
J032851	& 5.4		& 82.0	& 39.7	& 43.8	\\
J032859	& 18.5	& 30.0	& 49.7	& 61.4	\\
\hline
\end{tabular}
\begin{tablenotes}
\item[a] The uncertainty is estimated to be $\sim$30\% for the molecular abundances. 
\end{tablenotes}
\end{threeparttable}
\end{table}

\subsection{Dependence on bolometric luminosity}
\label{protostars}

\begin{figure}
  \centering           
     \includegraphics[width=2.8in]{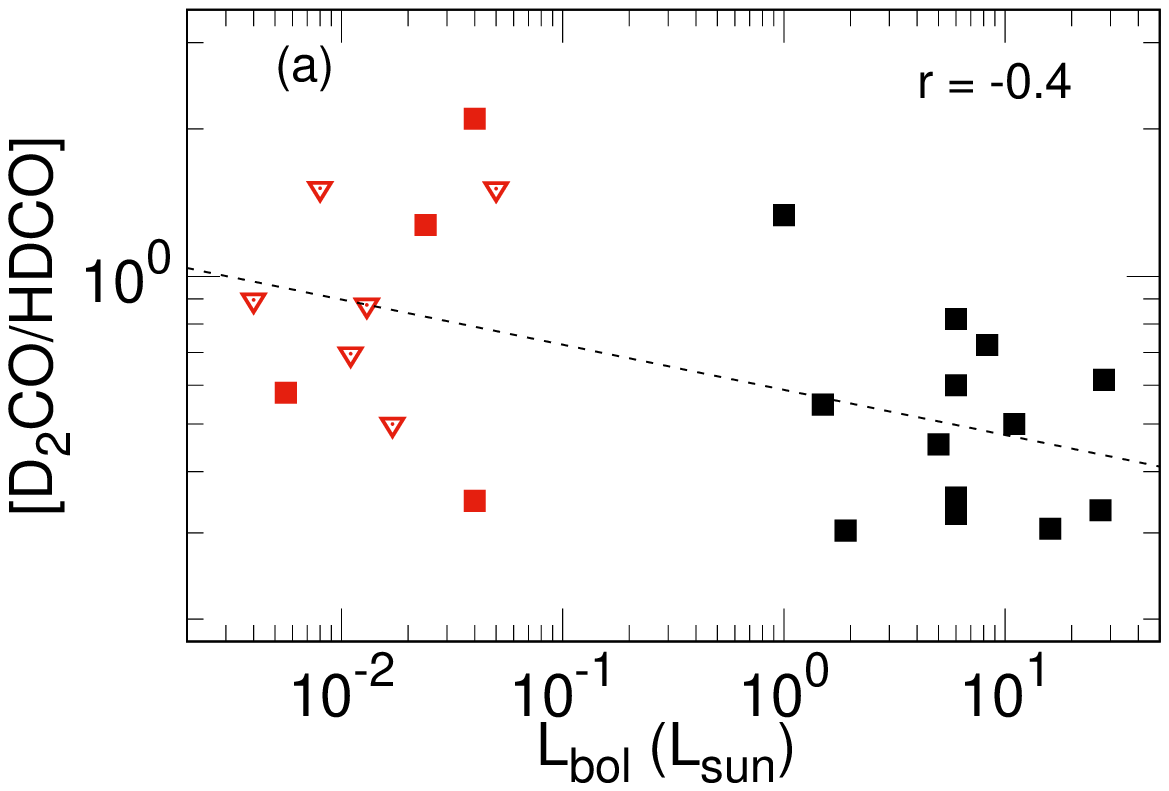}  
     \includegraphics[width=2.8in]{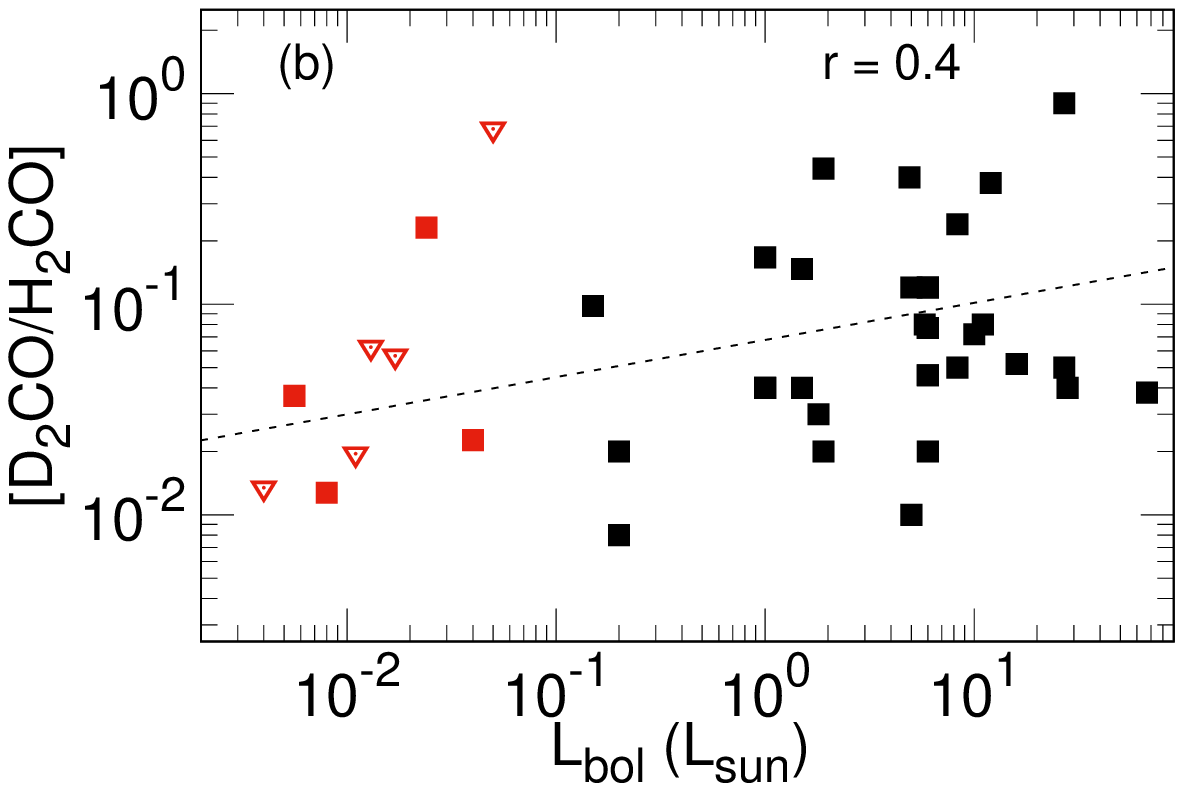}        
     \includegraphics[width=2.8in]{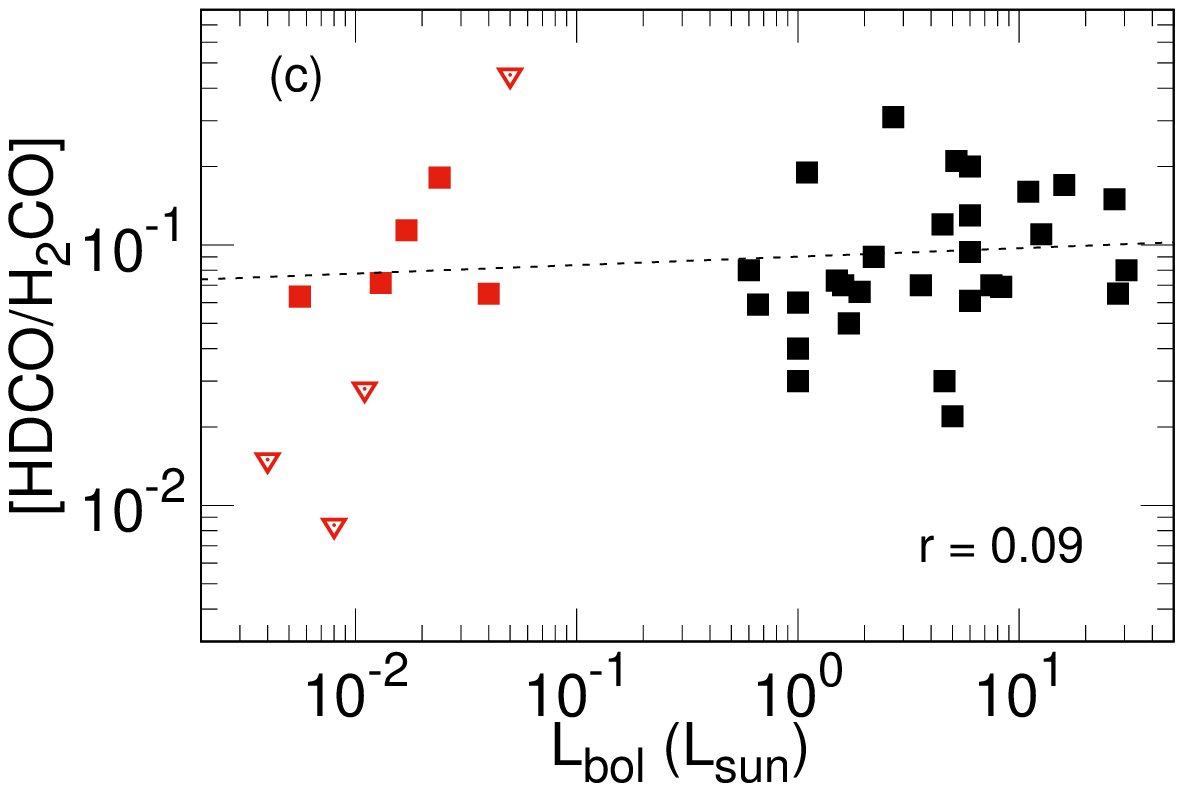}  
     \caption{Correlations between the D$_{2}$CO/HDCO ({\bf a}), D$_{2}$CO/H$_{2}$CO ({\bf b}), and HDCO/H$_{2}$CO ({\bf c}) ratios for the proto-BDs (red) and low-mass protostars (black). Diamonds denote the non-detections (upper limits). Dashed lines are the best-fit to the observed data points excluding the upper limits. The correlation coefficient for the best-fit is noted in the plot.  }
     \label{proto}     
  \end{figure}

Figure~\ref{proto} probes the trends in the D$_{2}$CO/HDCO, D$_{2}$CO/H$_{2}$CO, and HDCO/H$_{2}$CO ratios between Class 0/I proto-BDs and low-mass Class 0/I protostars. The data for the protostars has been compiled from the works of Roberts et al. (2002; 2007), Parise et al. (2006), Kang et al. (2015), Bacmann et al. (2003), and Loinard et al. (2002). Proto-BDs show comparatively higher D$_{2}$CO/HDCO but lower D$_{2}$CO/H$_{2}$CO ratios than protostars, with a similar correlation coefficient {\it r} $\sim$ 0.4 (Fig.~\ref{proto}ab). In contrast, the HDCO/H$_{2}$CO ratios are nearly constant with L$_{bol}$ (Fig.~\ref{proto}c). The main conclusion we can draw is that there is no significant change in these ratios across a wide range in L$_{bol}$ from $\sim$40 L$_{\sun}$ down to $\sim$0.002 L$_{\sun}$.

\subsection{Comparison with chemical models}
\label{models}

Since gas-phase chemistry is expected to produce exceedingly small level of multiple deuteration, active surface grain chemistry has been proposed for the formation of doubly-deuterated molecules such as D$_{2}$CO (e.g., Roberts \& Millar 2000ab; Loinard et al. 2002). Active grain surface chemistry proceeds efficiently by the deuteration of molecules accreted onto the grain surfaces (e.g., Tielens 1983). In the accretion model by Roberts \& Millar (2000ab), species colliding with dust grains are expected to stick, coating each grain in an icy mantle. The removal of species like CO, N$_{2}$, etc. from the gas-phase, which would normally destroy H$_{3}^{+}$ means that H$_{3}^{+}$ is more likely to react with HD and produce HD$^{+}$. This leads to a dramatic increase in deuterium fractionation of doubly deuterated species, with D/H ratios $>$0.01 and about 2 orders of magnitude higher than the ratios from pure gas-phase chemistry. Large enhancements of the molecular D/H ratios as observed here for proto-BDs provide evidence that accretion (grain surface chemistry) may be occurring for these molecules.

For a direct comparison, we consider the time-dependent gas-grain chemical model by Taquet et al. (2012; 2014) that maps the evolution of molecular deuteration from molecular clouds to protostellar envelopes. This model includes D and H addition, abstraction, and substitution surface reactions, and the spin states of H$_{2}$ and light ions. These models predict D$_{2}$CO/HDCO and HDCO/H$_{2}$CO ratios of $\sim$0.04 at 1$\times$10$^{5}$ yr that decrease by 1-2 orders of magnitude by 2$\times$10$^{5}$ yr. The predicted D$_{2}$CO/H$_{2}$CO ratios are 0.0016-0.0018 at 1$\times$10$^{5}$ yr that decrease by 2-3 orders of magnitude by 2$\times$10$^{5}$ yr. The model results show that including D and H abstraction and substitution reactions can produce high deuterium fractionation at moderate densities ($\sim$5$\times$10$^{6}$ cm$^{-3}$) and short timescale ($\sim$5000 yr) (Taquet et al. 2012; 2014). We also consider the gas-grain chemical model by Albertsson et al. (2013), which includes both gas-phase and surface species, and an extended multi-deuterated chemical network. These models predict D$_{2}$CO/HDCO and HDCO/H$_{2}$CO ratios in the range of $\sim$0.001-0.1 and D$_{2}$CO/H$_{2}$CO ratios in the range of $\sim$0.001- 0.01.

In comparison, the mean D$_{2}$CO/HDCO ratios of $\sim$1.1-1.2 for the proto-BDs are at least an order of magnitude higher than those predicted by these models. This is also noted for the mean D$_{2}$CO/H$_{2}$CO ratios of $\sim$0.05-0.1 for the proto-BDs that are at least a factor of $\sim$2 higher than the range predicted by the models. The mean HDCO/H$_{2}$CO ratios of $\sim$0.1-0.15 for the proto-BDs are comparable to the upper limit on these ratios predicted by the models. This emphasizes the need to include the physical conditions specific to proto-BDs in the chemical models with a deuterated network.

\section{Summary}

We have conducted IRAM 30-m observations in various HDCO and D$_{2}$CO transition lines for Class 0/I proto-BDs. Most of the detections are in the transition lines with the lowest upper energy level (E$_{upper} <$ 31 K) within our frequency range. None of the HDCO and D$_{2}$CO transition lines with E$_{upper} >$ 40 K between 213 and 279 GHz were detected for our proto-BD sample. The highest values are measured for the D$_{2}$CO/HDCO ratios, with a mean value of $\sim$1.1-1.25. The lowest are D$_{2}$CO/H$_{2}$CO ratios, with a mean value of $\sim$0.05-0.1. The D$_{2}$CO/HDCO ratios show a possible dependence on the evolutionary stage of the proto-BD, such that high ratios of $>$1 are only measured for the Stage 0 objects J182952 and J162625. We find a tentative trend of higher D$_{2}$CO/HDCO and lower D$_{2}$CO/H$_{2}$CO ratios in proto-BDs compared to low-mass protostars, while the HDCO/H$_{2}$CO ratios are nearly constant over L$_{bol}\sim$0.002--40 L$_{\sun}$. The mean D$_{2}$CO/HDCO, D$_{2}$CO/H$_{2}$CO, and HDCO/H$_{2}$CO ratios for the proto-BDs are comparatively higher than the range predicted by the current chemical models. Future dedicated modelling studies of the chemistry of proto-BDs can shed more light on the conditions under which deuterium fractionation is enhanced in these objects.

\section*{Acknowledgements}

B.R. acknowledges funding from the Deutsche Forschungsgemeinschaft (DFG) - Projekt number RI-2919/2-1. This work is based on observations carried out with the IRAM 30m telescope. IRAM is supported by INSU/CNRS (France), MPG (Germany) and IGN (Spain).

\section{Data Availability}

The data underlying this article are available in the IRAM archives through the VizieR online database.


\newpage

\title{Supplementary Material}

\appendix

\section{Spectra and Line Parameters}

\begin{figure*}
  \centering      
       \includegraphics[width=1.6in]{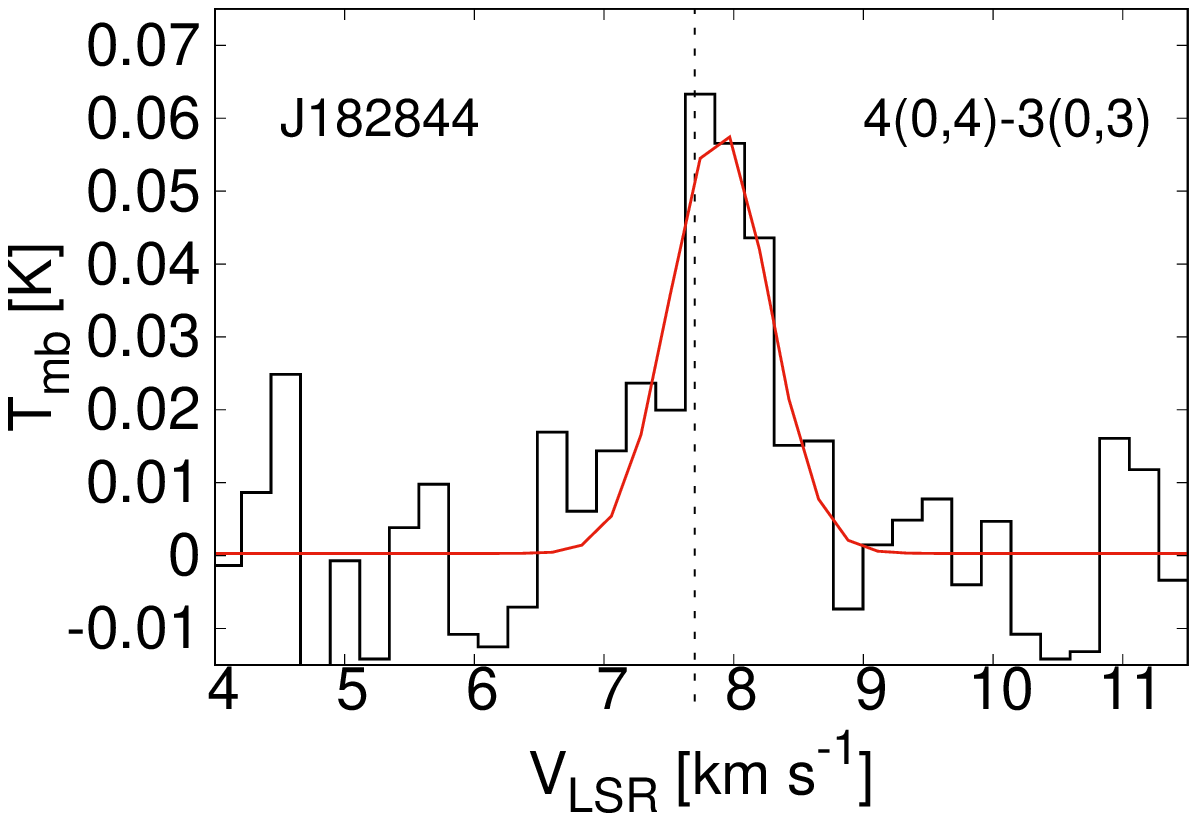}
       \includegraphics[width=1.6in]{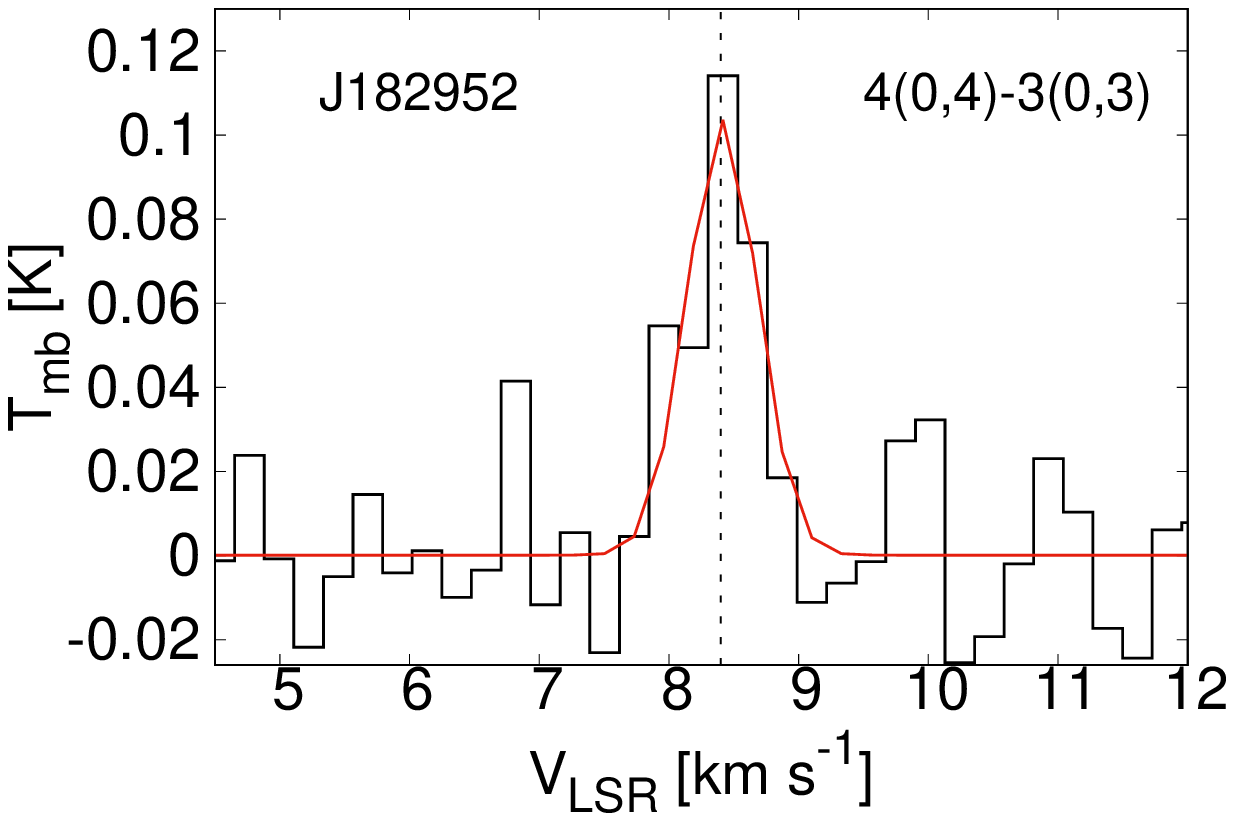}		  
       \includegraphics[width=1.6in]{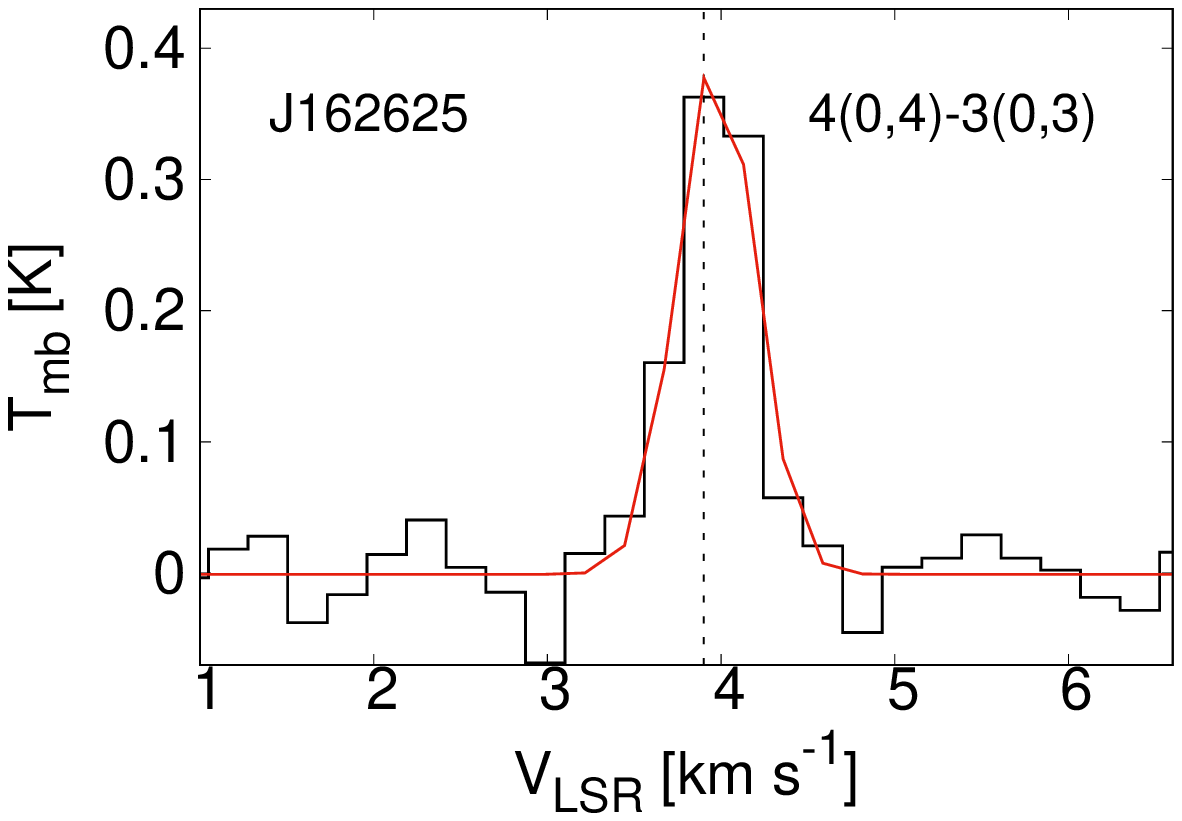}		
       \includegraphics[width=1.6in]{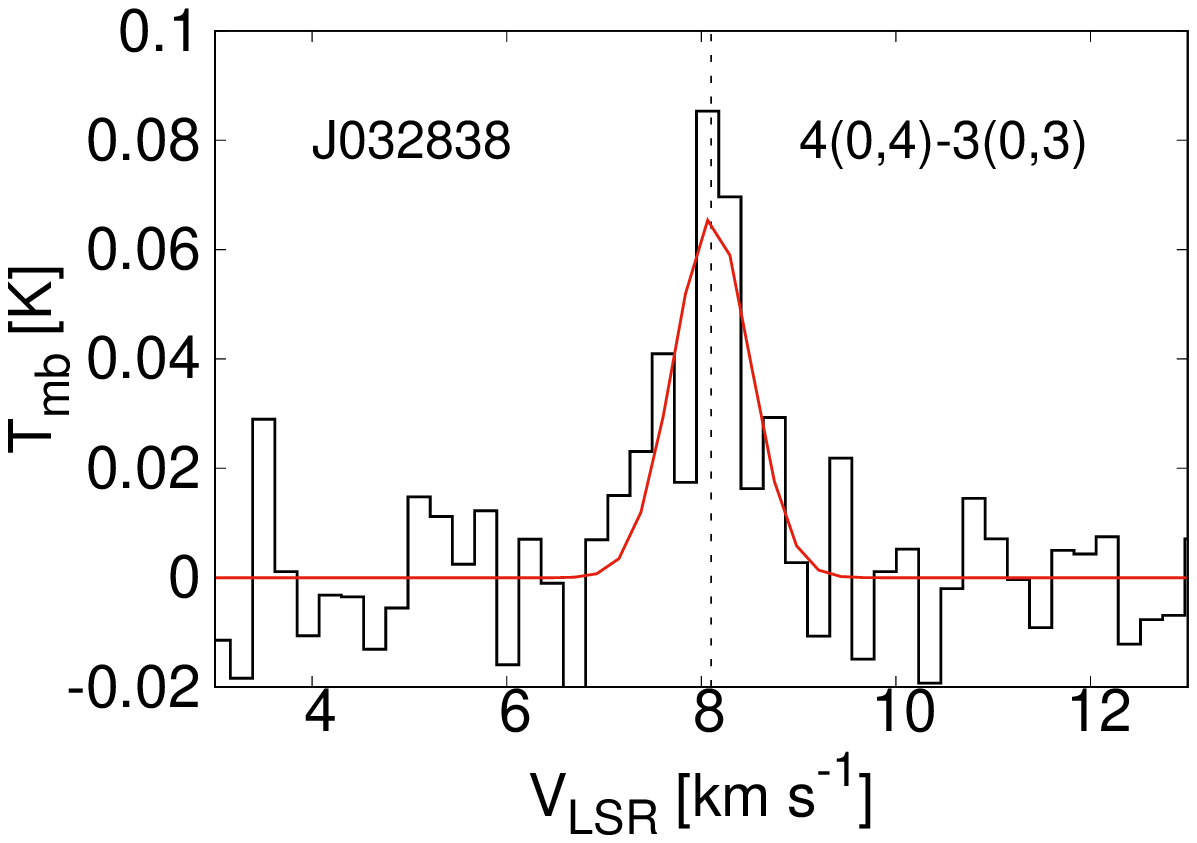}	
       \includegraphics[width=1.6in]{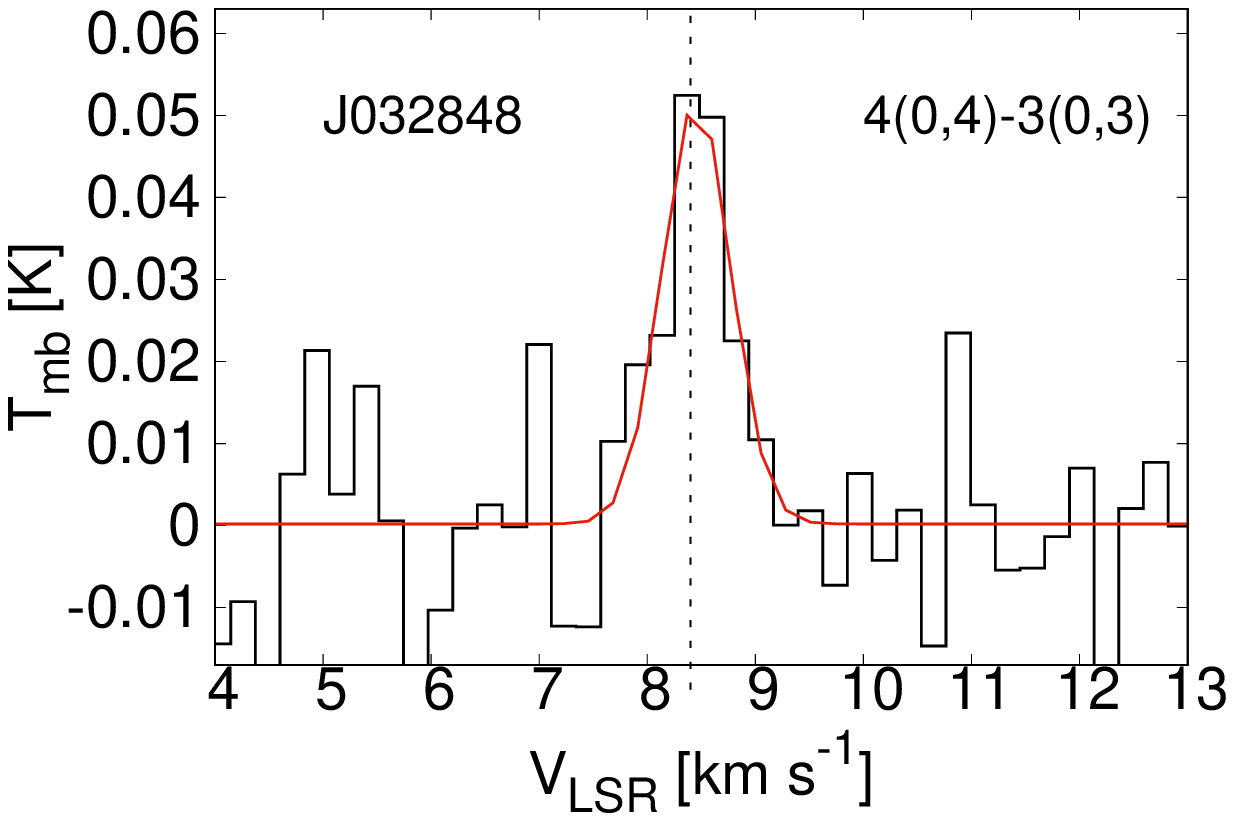}		       
       \includegraphics[width=1.6in]{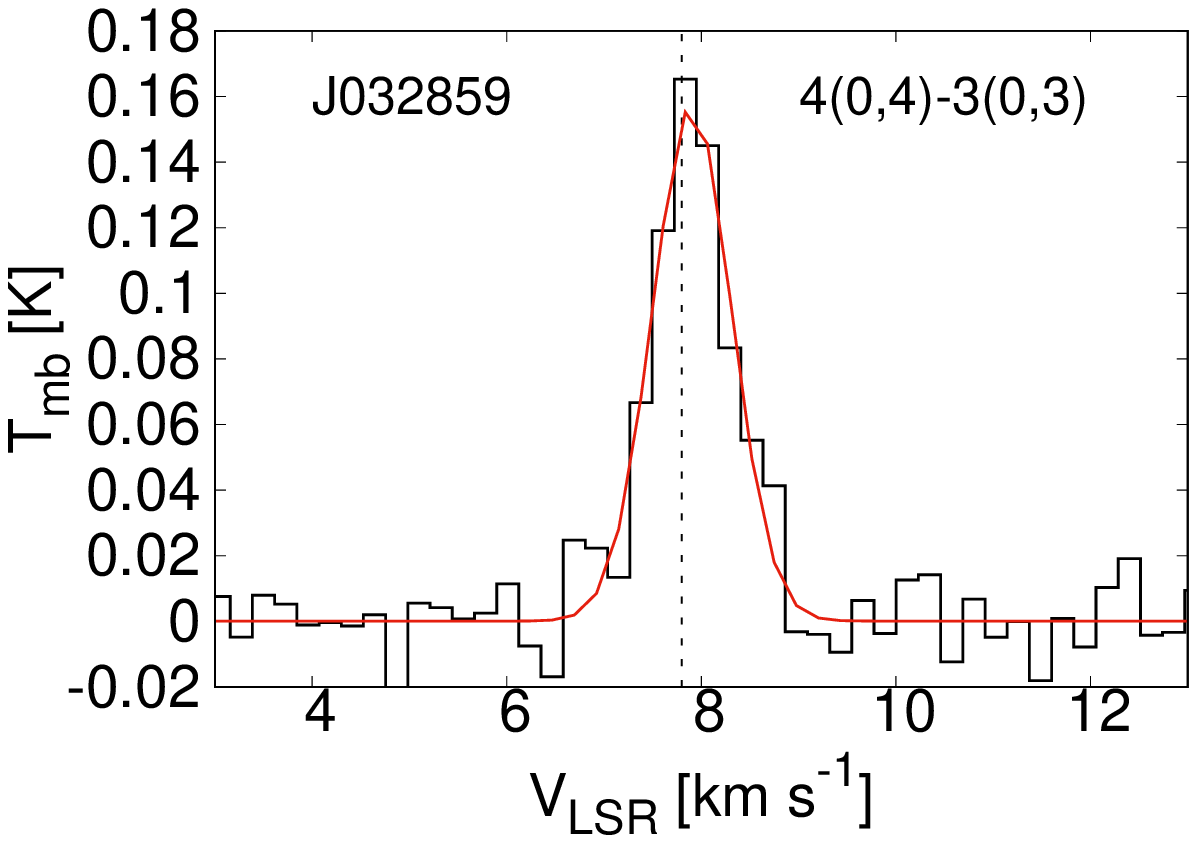}		
       \includegraphics[width=1.6in]{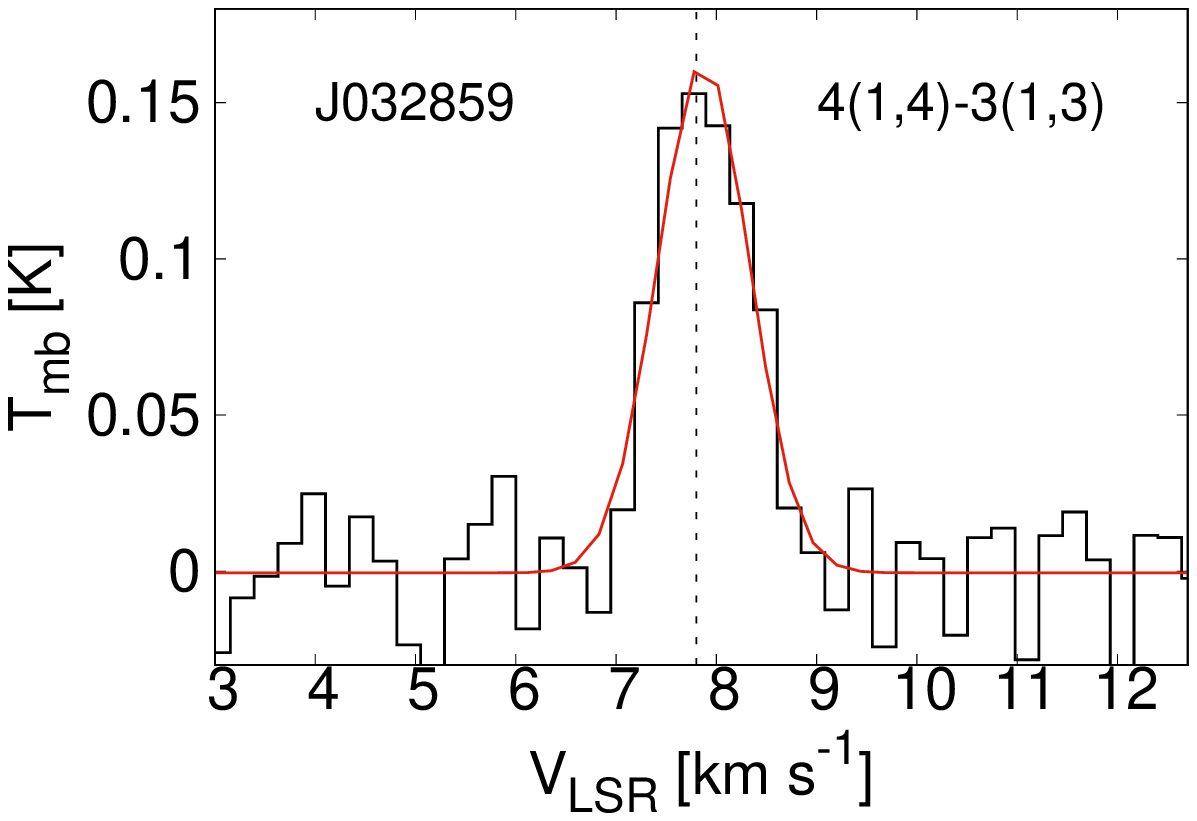}		 
       \includegraphics[width=1.6in]{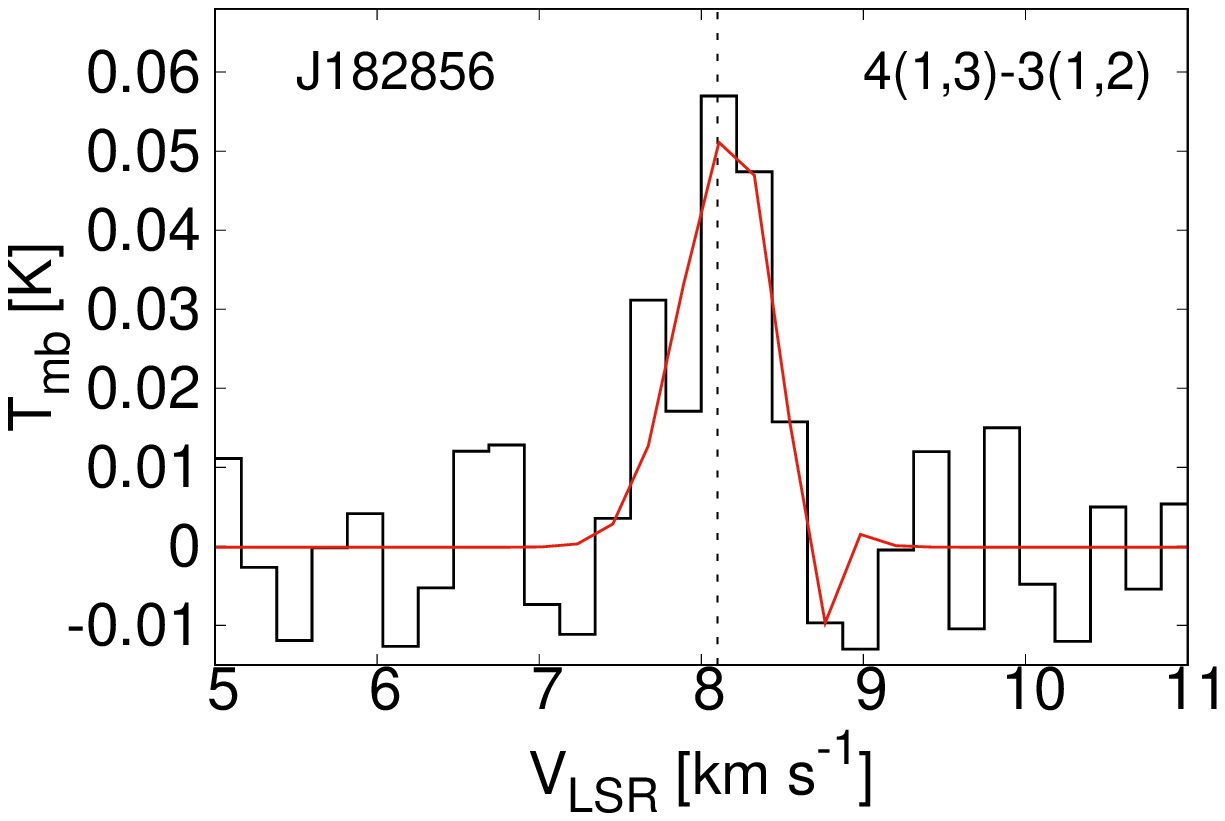}       
       \caption{The observed HDCO spectra (black) with the Gaussian fits (red). Black dashed line marks the V$_{lsr}$ listed in Table~\ref{hdco-pars}. }
     \label{hdco-figs}                
  \end{figure*}

\begin{figure*}
  \centering         
     \includegraphics[width=1.6in]{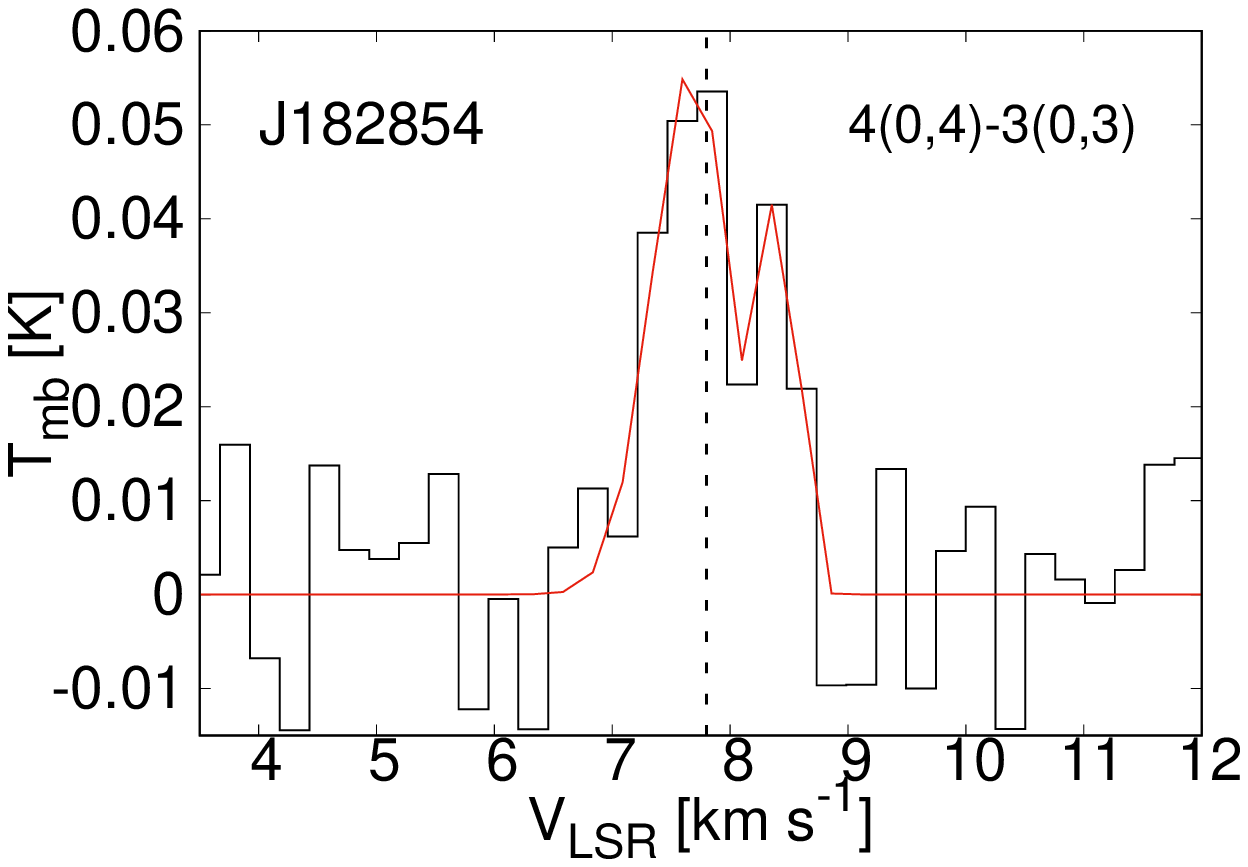}
     \includegraphics[width=1.6in]{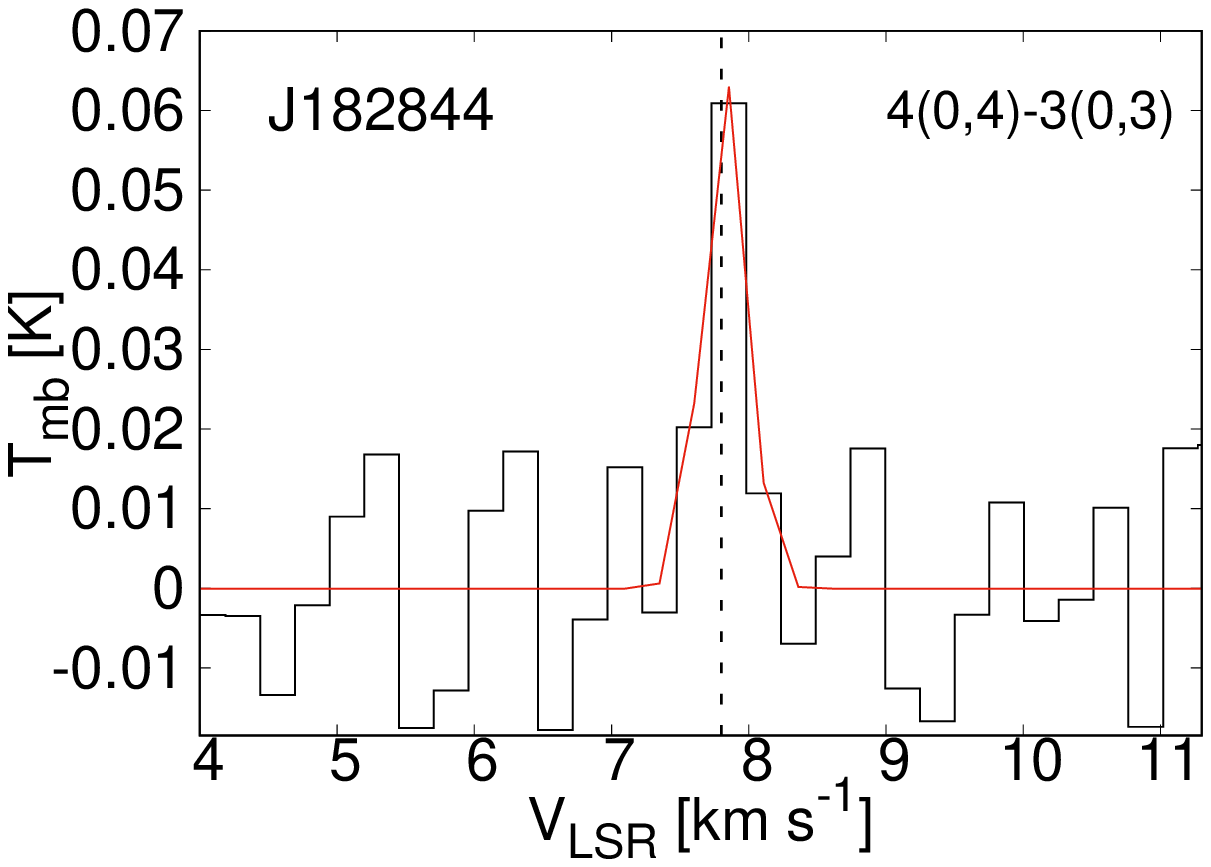}
     \includegraphics[width=1.6in]{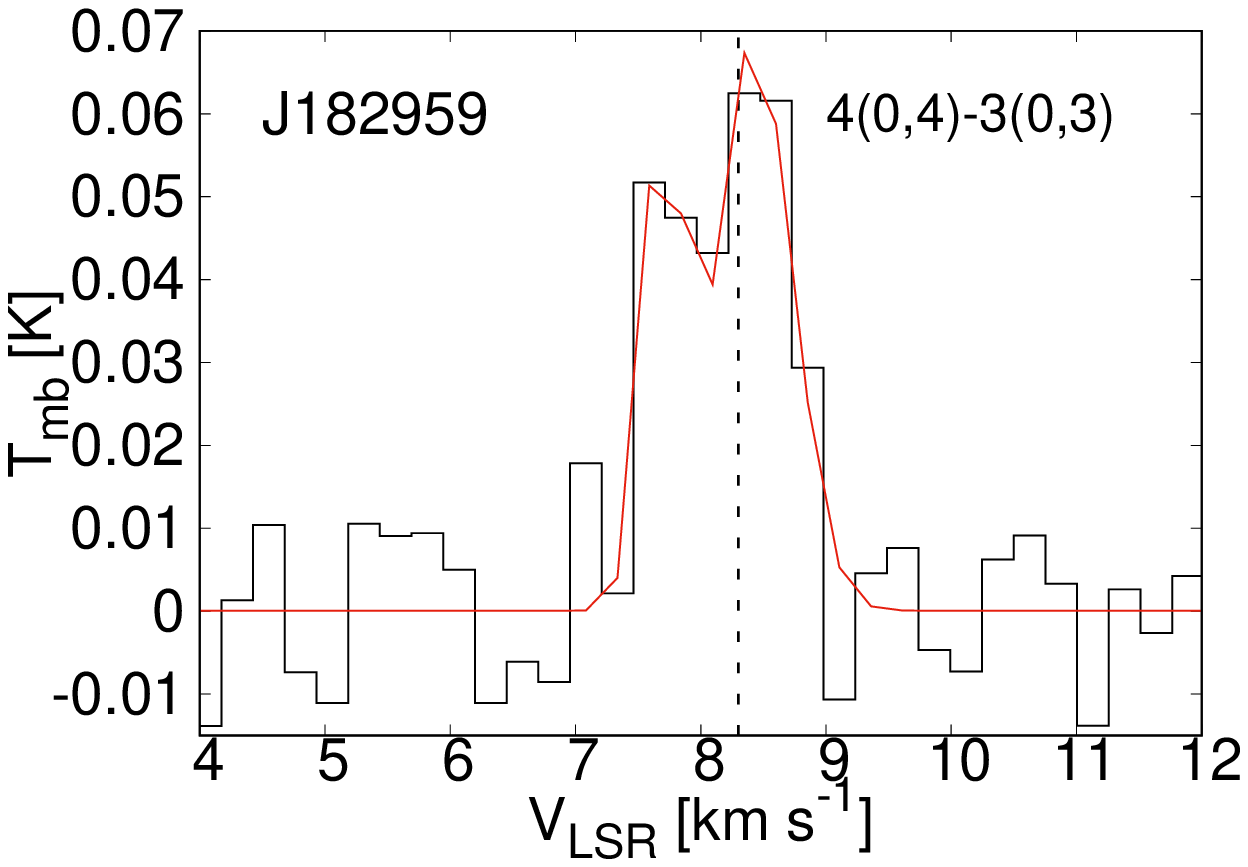}
     \includegraphics[width=1.6in]{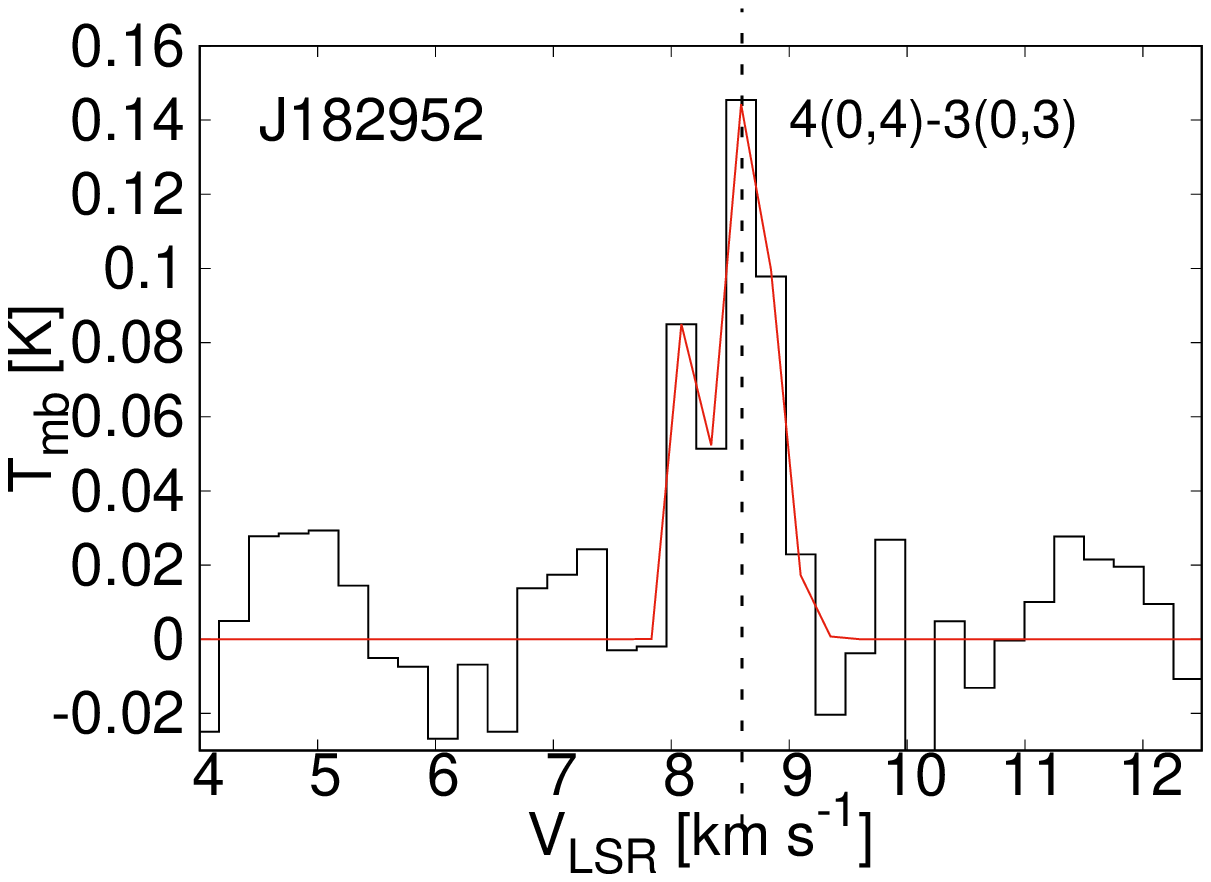}	
     \includegraphics[width=1.6in]{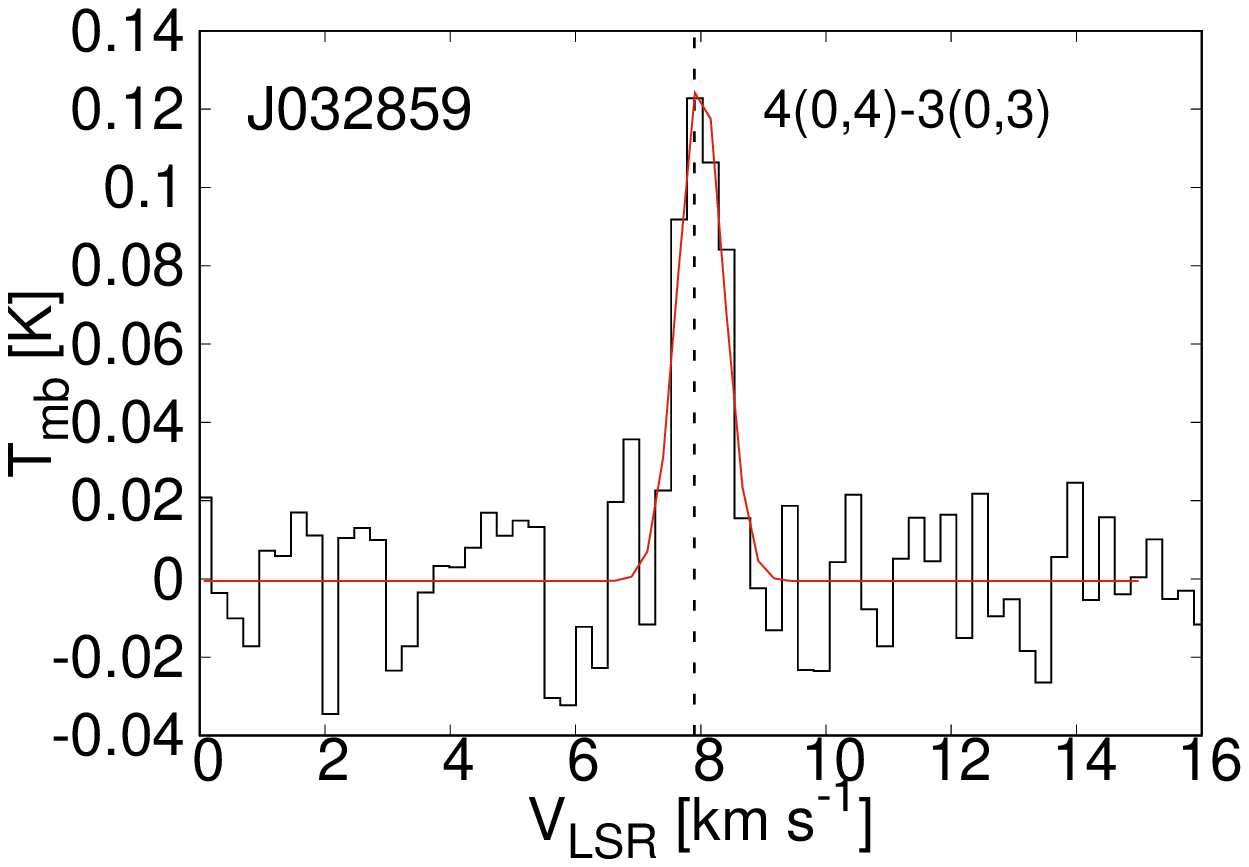}	   
     \includegraphics[width=1.6in]{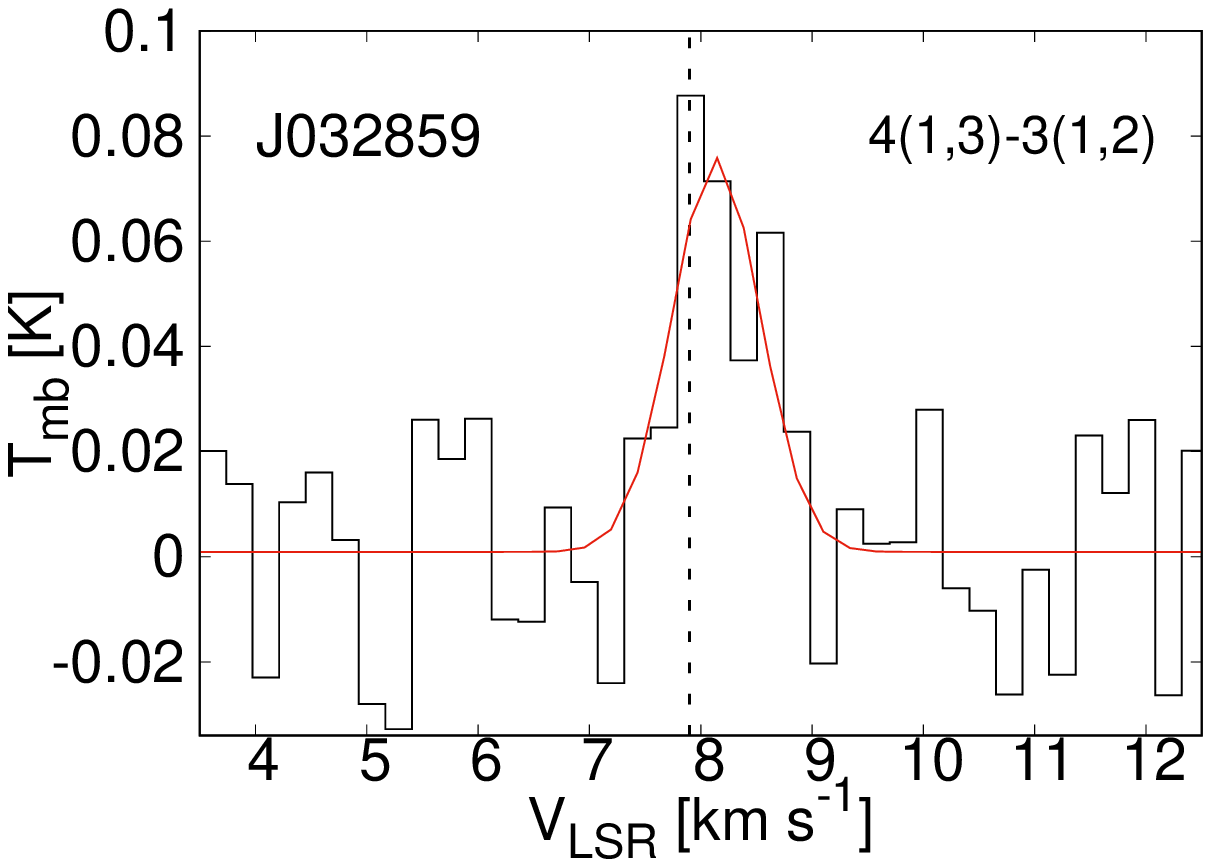}
     \includegraphics[width=1.6in]{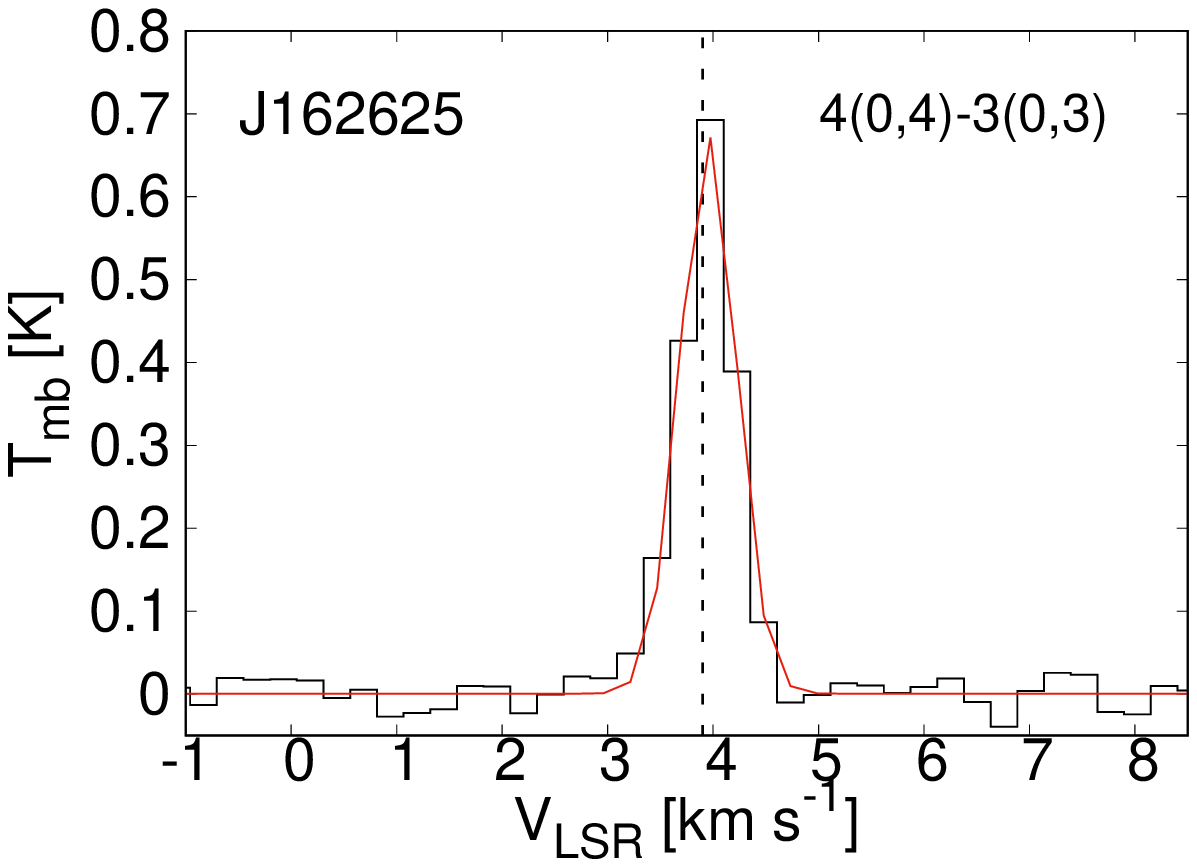}
     \includegraphics[width=1.6in]{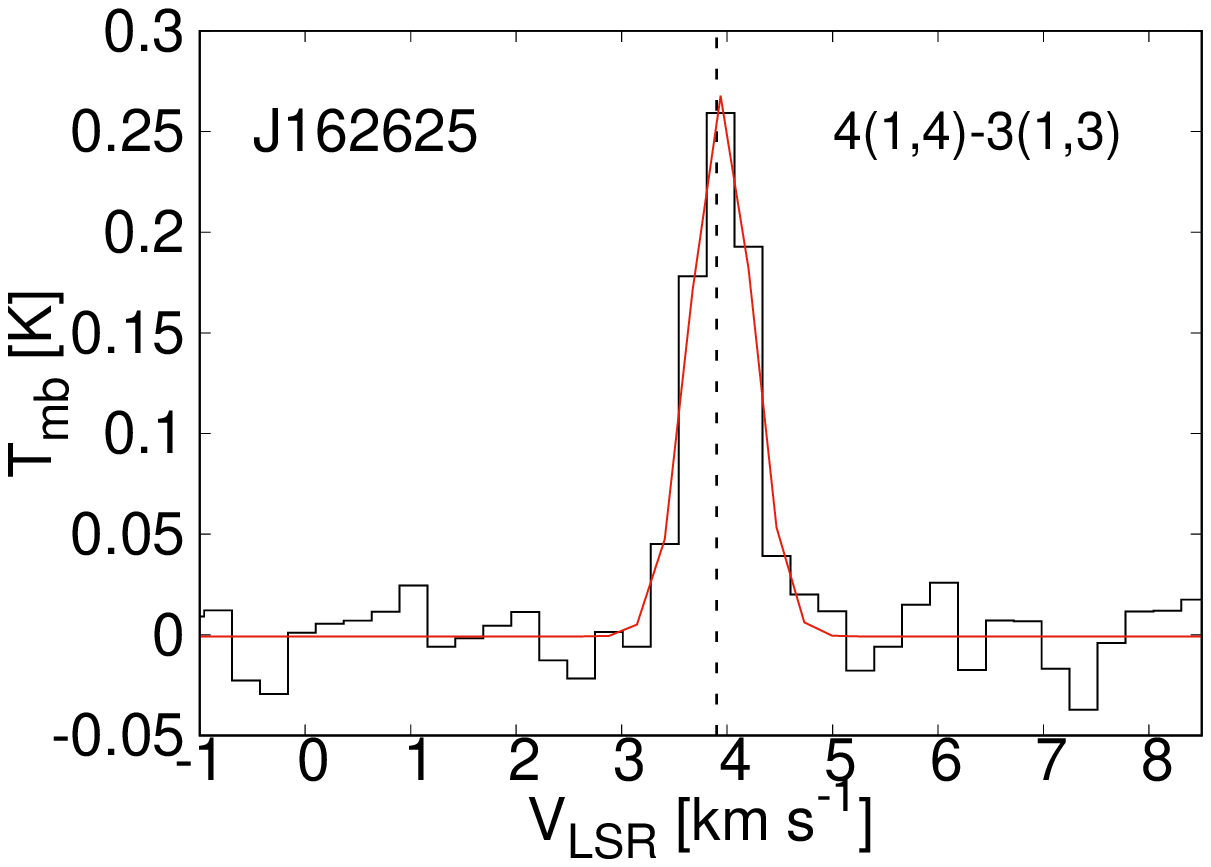}     
     \caption{The observed D$_{2}$CO spectra (black) with the Gaussian fits (red). Black dashed line marks the V$_{lsr}$ listed in Table~\ref{d2co-pars}.  }
     \label{d2co-figs}     
  \end{figure*}

The observed HDCO and D$_{2}$CO spectra are shown in Figs.~\ref{hdco-figs} and \ref{d2co-figs}, respectively. The line parameters derived from these spectra are listed in Tables~\ref{hdco-pars} and \ref{d2co-pars}.

\begin{table}
\centering
\caption{HDCO Line Parameters}
\label{hdco-pars}
\begin{threeparttable}
\begin{tabular}{ccccc} 
\hline
Object   	& V$_{lsr}$ 		& T$_{mb}$	& $\int{T_{mb} dv}$ 	&  $\Delta$v  \\
	     	& (km s$^{-1}$)		& (K)			& (K km s$^{-1}$) 	& (km s$^{-1}$)		\\
\hline
\multicolumn{5}{c}{o-HDCO 4(0,4)-3(0,3)}	\\ 
\hline 
J182854	&	8.0		&	<0.04	&	<0.04		& 	1.0	\\
J182844	&	7.7		&	0.06		&	0.05			&	0.8	\\ 	
J182959	&	8.0		&	<0.04	&	<0.04		& 	1.0	\\
J182856	&	8.0		&	<0.04	&	<0.04		& 	1.0	\\
J182952	&	8.4		&	0.11		&	0.07			&	0.6	\\	
J162625 	&	3.9		&	0.36		&	0.21			&	0.6	\\	
J032838	&	8.1		&	0.08		&	0.07			&	0.8	\\	
J032848	&	8.4		&	0.05		&	0.04			& 	0.8	\\   	
J032851	&	8.0		&	<0.04	&	<0.04		& 	1.0	\\
J032859	&	7.8 		&	0.16		&	0.16			&	1.0	 \\	
\hline
\multicolumn{5}{c}{p-HDCO 4(1,4)-3(1,3)}	\\ 
\hline 
J032859	&	7.8 		&	0.15		&	0.18			&	1.2	 \\	
\hline
\multicolumn{5}{c}{p-HDCO 4(1,3)-3(1,2)}	\\ 
\hline 
J182856	&	8.1		&	0.06		&	0.03			&	0.5	\\	
\hline
\end{tabular}
\end{threeparttable}
\end{table}

\begin{table}
\centering
\caption{D$_{2}$CO Line Parameters}
\label{d2co-pars}
\begin{threeparttable}
\begin{tabular}{llllllll} 
\hline
Object   	& V$_{lsr}$ 	& T$_{mb}$	& $\int{T_{mb} dv}$ 	&  $\Delta$v  \\
	     	& (km s$^{-1}$)	& (K)			& (K km s$^{-1}$) 	& (km s$^{-1}$)		\\
\hline
\multicolumn{5}{c}{o-D$_{2}$CO 4(0,4)-3(0,3)}	\\ 
\hline 
J182854 	& 7.8			& 0.05		& 0.07		& 1.3		\\
J182844 	& 7.8			& 0.07		& 0.02		& 0.3		\\
J182959 	& 8.3			& 0.06		& 0.07		& 1.2		\\
J182856	& 8.0			& <0.04		& <0.04		& 1.0		\\
J182952	& 8.6			& 0.13		& 0.10		& 0.7		\\ 	
J162625	& 3.9			& 0.7			& 0.5			& 0.6		\\	
J032838	& 8.0			& <0.04		& <0.04		& 1.0		\\
J032848	& 8.0			& <0.04		& <0.04		& 1.0		\\
J032851	& 8.4			& <0.05		& <0.03		& 0.6		\\	
J032859	& 7.9			& 0.12		& 0.10		& 0.8		\\	
\hline
\multicolumn{5}{c}{p-D$_{2}$CO 4(1,4)-3(1,3)}	\\ 
\hline 
J162625	& 3.9			& 0.19		& 0.11		& 0.6		\\
\hline
\multicolumn{5}{c}{p-D$_{2}$CO 4(1,3)-3(1,2)}	\\
\hline
J032859	& 7.9			& 0.05		& 0.05		& 1.0		\\	
\hline
\end{tabular}
\end{threeparttable}
\end{table}


\bsp	
\label{lastpage}
\end{document}